\documentclass[12pt,notitlepage]{iopart}
\usepackage{graphicx}
\usepackage{iopams,bbm,setstack}
\usepackage[utf8]{inputenc}
\usepackage[OT4]{fontenc}
\usepackage{url}
\usepackage{hyperref}
\usepackage{subcaption}
\newcommand{\Hil}{\mathcal{H}}
\newcommand{\Hvgr}{\Hil_{\rm vtx}^{\rm gr}}
\newcommand{\HVgr}{\Hil_{V}^{\rm gr}}

\newcommand{\HV}{\Hil_{V}}
\newcommand{\HVmat}{\Hil_{V}^{\rm mat}}
\newcommand{\C}{\hat{C}}
\newcommand{\Cgr}{\C^{\rm gr}}
\newcommand{\ket}[1][\Psi]{| #1 >}
\newcommand{\bra}[1][\Psi]{< #1 | }
\newcommand{\sign}{\mathsf{s}}
\newcommand{\braket}[2]{\left< #1 | #2 \right>}
\newcommand{\brakett}[1]{\left< #1 \right>}
\newcommand{\phys}[2]{\left< #1 | #2 \right>_{\rm phys}}
\newcommand{\physp}[2]{\left< #1 | #2 \right>_{+,\epsilon}}
\newcommand{\physm}[2]{\left< #1 | #2 \right>_{-,\epsilon}}
\newcommand{\physpm}[2]{\left< #1 | #2 \right>_{\pm,\epsilon}}

\newcommand{\Texp}{{\rm Texp}}
\newcommand{\Nspace}{\mathcal{N}}
\newcommand{\Hphys}{\Hil_{\rm phys}}
\newcommand{\iu}{{\it i}}
\newcommand{\norm}[1]{||#1||}

\newcommand{\I}{\mathbbm{1}}

\newcommand{\p}{\hat{p}_{\phi}}
\newcommand{\ep}{p_{\phi}}
\newcommand{\Ep}{E_{\phi}}
\newcommand{\Res}{\rm Res}
\newcommand{\sgn}{\rm sgn}
\newcommand{\slice}{\Sigma}
\newcommand{\supp}{{\rm supp}}
\newcommand{\Diff}{{\rm Diff}}
\newcommand{\DV}{\Diff_V}
\newcommand{\inn}{{\rm in}}
\newcommand{\out}{{\rm out}}
\newcommand{\Bohr}[1][\varphi]{d\mu_{\rm Bohr}\left(#1\right)}
\newcommand{\J}{\hat{J}}
\newcommand{\link}{\ell}
\newcommand{\loops}{\mathcal{L}}
\newcommand{\spacetime}{\mathcal{M}}
\newcommand{\interval}{I}
\newcommand{\complex}{\kappa}
\newcommand{\Amplitude}[1]{\mathcal{A}_{#1}}

\newcommand{\Inv}[1]{{\rm Inv}\left( #1 \right)}
\newcommand{\R}{\mathcal{R}}
\newcommand{\node}{n}
\newcommand{\omicron}{{\cal I}}
\newcommand{\contractor}{\mathcal{A}}
\newcommand{\TDiff}[1]{{\rm TDiff}_{#1}}
\newcommand{\nodes}[1]{{\rm Nodes}(#1)}
\newcommand{\links}[1]{{\rm Links}(#1)}
\newcommand{\NV}{|V |!}

\newcommand{\spinfoam}{F}

\usepackage{accents}
\newcommand{\dbtilde}[1]{\accentset{\approx}{#1}}
\def\be{\begin{equation}}
\def\ee{\end{equation}}
\def\eqref{\eref}
\begin{document}
\title[Spin foams with scalar field]{Spin-foam model for gravity coupled to massless scalar field}
\author{Marcin Kisielowski$^{1,2}$, Jerzy Lewandowski$^1$}
\address{${}^1$ Instytut Fizyki Teoretycznej, Uniwersytet Warszawski,
ul. Pasteura 5, 02-093 Warsaw, Poland}
\address{${}^2$ Institute for Quantum Gravity, Chair for Theoretical Physics III, University of Erlangen-Nürnberg, Staudtstraße 7 / B2,
91058 Erlangen, Germany}
\ead{Marcin.Kisielowski@fuw.edu.pl, Jerzy.Lewandowski@fuw.edu.pl}
\begin{abstract}
A spin-foam model is derived from the canonical model of Loop Quantum Gravity coupled to a massless scalar field. We generalized to the full theory the scheme first proposed in the context of Loop Quantum Cosmology by Ashtekar, Campiglia and Henderson, later developed by Henderson, Rovelli, Vidotto and Wilson-Ewing.
\end{abstract}
\maketitle
\section{Introduction}
Our motivation to study the coupling of spin foams with the massless scalar field comes from several directions. The first direction is the spin-foam theory \cite{Zakopane,LQG25,Baezintro,PerezOldReview,PerezNewReview,EngleReview,AshtekarRovelliReuterRev,rovelli2014covariant}. The research in the theory has been focused on studying gravitational field and there are interesting proposals for such models \cite{BC,EPRL,FK,BaratinOritiI,BaratinOritiII}. The idea behind such approach is to derive first a well-tested model of gravity and couple it to gravity at a later point. However, in principle, the theory of gravity interacting with matter fields can be quite different from the theory of pure gravity and it may turn out to be simpler \cite{BarrettMatter}. In fact, a simplification has been observed in the canonical Loop Quantum Gravity where the scalar field becomes a convenient "clock" solving the problem of time and reducing the problem of finding the physical scalar product to the problem of finding matrix elements of an evolution operator. This brings us to the second direction: canonical Loop Quantum Gravity \cite{StatusReport,ThiemannBook,RovelliBook,Zakopane,LQG25,MaReview,AshtekarRovelliReuterRev}. Spin foams are believed to be related to the canonical Loop Quantum Gravity. Although some relation has been found \cite{RR, EPRL, SFLQG}, the precise link was missing. Such link could be technically very useful: spin foams are believed to provide an approximate expansion, called the vertex expansion, of the physical scalar product. The matrix elements of the constraint operators are well known but finding the spectrum and (generalized) eigenvectors of the scalar constraint operator are still an open problem, which is probably the main technical obstruction for modelling the physical processes. This brings us to the third direction. Physical predictions from Loop Quantum Gravity are derived within Loop Quantum Cosmology, which is based on very simplified models where the degrees of freedom are drastically reduced at the classical level and later quantized. An open problem is to study the physical processes within the full theory, where the reduction of the degrees of freedom is done at the quantum level or only approximate. An interesting approach to this problem has been proposed by the Marseille's group \cite{BRV} opening new discipline called spin-foam cosmology. However, it relies on a spin-foam model of pure gravity and uses a vertex expansion which is motivated by an incomplete relation between the spin-foam theory and the canonical Loop Quantum Gravity. We hope that the research reported in this paper will provide the missing ingredients and will become the next step towards bringing the theory closer to observations.

We propose a spin-foam model of Quantum Gravity coupled with a massless scalar field. We derive the model from the canonical model proposed by Domagała, Dziendzikowski and Lewandowski \cite{DziendzikowskiDomagalaLewandowski} (see also \cite{GravityQuantized}) providing this way a precise link between the canonical Loop Quantum Gravity and covariant spin-foam theory. We use the version of the gravitational scalar constraint from \cite{NewScalarConstraint, HamiltonianOperator}.  Our derivation is based on a similar computation by Ashtekar, Campiglia and Henderson \cite{AshtekarSpinFoamI,AshtekarSpinFoamII,AshtekarSpinFoamIII} in Loop Quantum Cosmology. The expansion they found is non-local. Therefore one cannot expect that direct generalization of their ideas to the full theory could lead to a proper spin foam model. This problem was solved by Henderson, Rovelli, Vidotto and Wilson-Ewing \cite{Localsfexpansion} by introducing a regulator. Still, the authors considered pure gravity only and after removing the regulator an expression was found in terms of Dirac delta and its derivatives. Since, it is not known if in the full theory the gravitational part of the scalar constraint operator has continuous spectrum at $0$, it is not known if a generalization of this expression would be well defined. Therefore we decided to focus on a theory coupled to a scalar field, where there is no such issue, because the spectrum of the momentum operator is the whole real line. In section \ref{sc:LQC} we describe the derivation in Loop Quantum Cosmology. We recall the basic ideas from the papers \cite{AshtekarSpinFoamI,AshtekarSpinFoamII,AshtekarSpinFoamIII, Localsfexpansion} and use the regulator from \cite{Localsfexpansion} to find a local expansion of the physical scalar product in the model with massless scalar field. We show that the formulas from \cite{AshtekarSpinFoamI,AshtekarSpinFoamII,AshtekarSpinFoamIII} are recovered after the regulator is removed. Our derivation is based on new operator approach. This approach is generalized in the next section \ref{sc:LQG} to the full theory and leads first to operator spin foams where the edges are colored with certain operators and vertices are colored with contractors (see also \cite{OSM,Feynman,KisielowskiPhD}). It is transformed afterwards into standard formulation in terms of spin-foam amplitudes. An important technical obstacle, which we needed to overcome, was that in the known models either the scalar field operator or the scalar field momentum operator is well defined. We chose the polymer representation where the momentum operator is well defined. The price that we paid was that the expected sum over the momenta had to be replaced by a Lebesgue integral over the momenta. Finally, in section \ref{sc:Example} we provide a basic example illustrating how our construction works in practice.
 
\newcommand{\CE}[1][ ]{\hat{K}_{#1}}
\newcommand{\eCE}[1][ ]{K_{#1}}
\newcommand{\CL}[1][ ]{\hat{D}_{#1}}
\newcommand{\eCL}[1][ ]{D_{#1}}
\section{Spin-foam paradigm in Loop Quantum Cosmology}\label{sc:LQC}
\subsection{Pure gravity}
\subsubsection{ACH approach with HRVW regulator}\label{sc:ACHHRVW}
In their work Ashtekar, Campiglia and Henderson proposed a spin-foam formulation of Loop Quantum Cosmology \cite{AshtekarSpinFoamI,AshtekarSpinFoamII,AshtekarSpinFoamIII}. First, they found a perturbative expression for the matrix elements of the evolution operator generated by the gravitational Hamiltonian constraint operator $\alpha \Cgr$
$$
\braket{\nu_f}{e^{-\iu \alpha \Cgr} \nu_i},
$$
where $\alpha\in \mathbb{R}$. We consider a basis $\ket[\nu]$ of the Hilbert space $\Hil$ on which the operator $\Cgr$ is (densely) defined. The operator $\Cgr$ is split into its diagonal part $\CL$ and off-diagonal part $\CE$ in this basis:
$$
\Cgr=\CL+\CE.
$$
In this notation the vectors $\ket[\nu]$ are eigenvectors of the operator $\CL$. The eigenvalues of $\CL$ will be denoted by $\CL[\nu \nu]$:
$$\eCL[\nu \nu]=\braket{\nu}{\CL \nu}.$$
The matrix elements of $\CE$ will be denoted by $\CE[\mu \nu]$:
$$
\eCE[\mu \nu]=\braket{\mu}{\CE \nu}.
$$
The perturbative expansion is in the number of actions of the off-diagonal part $\CE$. Ashtekar, Campiglia and Henderson introduced an auxiliary expansion parameter $\lambda$ and considered evolution generated by the operator $\alpha \Cgr_{\lambda}$, where 
$$
\Cgr_{\lambda}=\CL+\lambda \CE.
$$
At the end the matrix elements of the evolution operator are evaluated at $\lambda=1$. The expansion is obtained by using the interaction picture. The interaction Hamiltonian is
$$
H_I(\tau)=e^{\iu \alpha \CL \tau} \alpha \CE e^{-\iu \alpha \CL \tau}.
$$
The evolution in the interaction picture is described by
$$
\tilde{U}_{\lambda}(\tau)=e^{\iu \alpha \CL \tau} e^{-\iu \alpha \Cgr_\lambda}.
$$
Since $\tilde{U}_{\lambda}(\tau)$ satisfies the differential equation
$$
\frac{d \tilde{U}_{\lambda}}{d\tau}(\tau)=-\iu \lambda H_{I}(\tau)  \tilde{U}_{\lambda}(\tau),
$$
it is of the form
\begin{eqnarray*}
 \tilde{U}_{\lambda}(\tau)=\Texp(-\iu \lambda \int_0^{\tau} H_{I}(\tau)d \tau)=\\=\I+\sum_{M=1}^{\infty} \lambda^M \int_{0}^{\tau}d \tau_M\int_{0}^{\tau_M}d \tau_{M-1}\ldots \int_{0}^{\tau_2} d\tau_1  (-\iu H_I(\tau_M))\ldots (-\iu H_I(\tau_1)).
\end{eqnarray*}
Using this expansion Ashtekar, Campiglia and Henderson wrote the matrix elements of the evolution operator in the following form:
$$
\braket{\nu_\out}{ e^{-\iu \alpha \Cgr_\lambda}\nu_\inn}=\braket{\nu_\out}{e^{-\iu \alpha \CL} \tilde{U}_{\lambda}(1) \nu_\inn}=\braket{\nu_\out}{ e^{-\iu \alpha \CL}\nu_\inn}+\sum_{M=1}^{\infty}\lambda^M\sum_{\nu_1,\ldots,\nu_{M-1}}  A(\nu_M,\ldots, \nu_0, \alpha),
$$
where $\nu_0=\nu_\inn$, $\nu_M=\nu_\out$,
$$
A(\nu_M,\ldots, \nu_0, \alpha):=\int_{0}^{1}d \tau_M\ldots \int_{0}^{\tau_2} d\tau_1 e^{-\iu (1-\tau_M)\eCL[\nu_M\nu_M]}(-\iu \alpha \eCE[\nu_M \nu_{M-1}])\ldots (-\iu \alpha \eCE[\nu_1 \nu_0]) e^{-\iu \alpha \tau_1 \eCL[\nu_0\nu_0]},
$$
They calculated the integrals over $\tau_1,\ldots,\tau_M$ and obtained the formula for $A(\nu_M,\ldots, \nu_0, \alpha)$:
$$
A(\nu_M,\ldots, \nu_0, \alpha)=\eCE[\nu_M \nu_{M-1}]\ldots \CE[\nu_1 \nu_{0}] \prod_{k=1}^p \frac{1}{(n_k-1)!}(\frac{\partial}{\partial \eCL[w_k w_k]})^{n_k-1}\sum_{m=1}^p \frac{e^{-\iu \alpha \eCL[w_m w_m]}}{\prod_{j\neq m}^p(\eCL[w_m w_m]-\eCL[w_j w_j])},
$$
where following \cite{AshtekarSpinFoamII} we label by $\eCL[w_m w_m]$ the $p$ distinct values in the sequence $(\eCL[\nu_M \nu_M],\ldots,\eCL[\nu_0 \nu_0])$ and by $n_m$ the number of times $\eCL[w_m]$ appears in this sequence. Clearly, $n_m$ satisfy $n_1+\ldots +n_p=M+1$.

We will be interested in the physical scalar product, which can be obtained by a group averaging procedure:
$$
\phys{\nu_f}{\nu_i}=\frac{1}{2\pi}\int_{-\infty}^{\infty} d\alpha\braket{\nu_f}{ e^{-\iu \alpha \Cgr}\nu_i}.
$$
As noted by Henderson, Rovelli, Vidotto and Wilson-Ewing \cite{Localsfexpansion} this integral leads to a formal expression:
$$
\phys{\nu_\out}{\nu_\inn} =\braket{\nu_\out}{\delta(\CL)\,\nu_\inn}+\sum_{M=1}^{\infty}\lambda^M\sum_{\nu_1,\ldots,\nu_{M-1}}  A(\nu_M,\ldots, \nu_0),
$$
where
$$
A(\nu_M,\ldots, \nu_0)=\frac{1}{2\pi}\int_{-\infty}^\infty A(\nu_M,\ldots, \nu_0,\alpha)
$$
can be expressed in terms of Dirac deltas and it's derivatives:
$$
A(\nu_M,\ldots, \nu_0)=\eCE[\nu_M \nu_{M-1}]\ldots \eCE[\nu_1 \nu_{0}] \prod_{k=1}^p \frac{1}{(n_k-1)!}(\frac{\partial}{\partial \eCL[w_k w_k]})^{n_k-1}\sum_{m=1}^p \frac{\delta(\eCL[w_m w_m])}{\prod_{j\neq m}^p(\eCL[w_m w_m]-\eCL[w_j w_j])}.
$$
They note that this expression does not have a local form. As a result it is not of the form of standard spin-foam amplitudes. They introduce a regulator which brings this expression into a local form:
$$
A_\epsilon(\nu_M,\ldots, \nu_0):=\frac{1}{2\pi}\int_{-\infty}^\infty A(\nu_M,\ldots, \nu_0,\alpha)e^{-\epsilon |\alpha|}= A_{-\epsilon}(\nu_M,\ldots, \nu_0) + A_{+\epsilon}(\nu_M,\ldots, \nu_0),
$$
were 
$$
A_{-\epsilon}(\nu_M,\ldots, \nu_0):=\frac{1}{2\pi} \int_0^{\infty } A(\nu_M,\ldots, \nu_0,\alpha)e^{-\epsilon \alpha},
$$
$$
A_{+\epsilon}(\nu_M,\ldots, \nu_0):=\frac{1}{2\pi} \int_{-\infty }^0 A(\nu_M,\ldots, \nu_0,\alpha)e^{\epsilon \alpha}.
$$
After performing the integrations they obtain:
$$
A_{+\epsilon}(\nu_M,\ldots, \nu_0)=\frac{1}{2\pi} \eCE[\nu_M \nu_{M-1}]\ldots \eCE[\nu_1 \nu_{0}] \frac{\iu (-1)^M}{ \prod_{m=0}^M (\eCL[\nu_m \nu_m] +\iu \epsilon)}.$$
Let us note that the sums
$$
\sum_{\nu_1,\ldots, \nu_{M-1}} A_{+\epsilon}(\nu_M,\ldots, \nu_0)
$$ 
can be written as a matrix elements of an operator
$$
\sum_{\nu_1,\ldots, \nu_{M-1}} A_{+\epsilon}(\nu_M,\ldots, \nu_0) =- \frac{(-1)^M}{2\pi \iu} \braket{\nu_M}{(\CL+\iu \epsilon)^{-1} (\CE (\CL+\iu\epsilon)^{-1})^M\nu_0}.
$$
Similarly, an operator for the sum involving $A_{-\epsilon}(\nu_M,\ldots, \nu_0)$ can be found. We will rederive this formula in the next subsection using an operator approach.
\subsubsection{Operator approach}
\renewcommand{\CE}[1][ ]{\hat{V}_{#1}}
\renewcommand{\CL}[1][0]{\hat{H}_{#1}}
Let $\CL$ and $\lambda\CE$ be the diagonal and off-diagonal parts of a bounded self-adjoint operator $\C_\lambda$ (not necessarily the gravitational constraint operator) in some basis:
$$\C_\lambda=\CL+\lambda \CE.$$ Let us study quantum dynamics of a system with $\C_\lambda$ as quantum constraint operator. Our goal is to find a Hilbert space of solutions of the constraint. This can be achieved by defining an expression of the form:
\be\label{eq:phys}
"\phys{\Psi_\out}{\Psi_\inn}=\bra[\Psi_\out] \delta(\C_\lambda) \ket[\Psi_\inn]."
\ee
It will be called a physical scalar product. Let $\Nspace$ be the null space, i.e. the space of vectors $\Psi$ for which
$$
\phys{\Psi}{\Psi}=0.
$$
The Hilbert space of solutions of the constraint is the quotient of the 
$$
\Hphys=\Hil\slash \Nspace.
$$

Let us note, that the Dirac delta $\delta$ can be expressed as a limit
$$
\delta(x)=\lim_{\epsilon\to 0} \frac{1}{\pi}\frac{\epsilon}{x^2+\epsilon^2}=\frac{1}{2\pi\iu}\lim_{\epsilon\to 0}\left(\frac{1}{x-\iu \epsilon}-\frac{1}{x+\iu \epsilon}\right).
$$
As a result we will define \eqref{eq:phys} as a limit
$$
\phys{\Psi_\out}{\Psi_\inn}:=\frac{1}{2\pi \iu}\lim_{\epsilon\to 0}\left( \bra[\Psi_\out] (\C_\lambda-\iu \epsilon)^{-1} \ket[\Psi_\inn]-\bra[\Psi_\out] (\C_\lambda+\iu \epsilon)^{-1} \ket[\Psi_\inn]\right).
$$
In the following we will adopt the notation from \cite{Localsfexpansion} and write
$$
\physp{\Psi_\out}{\Psi_\inn}:= \frac{1}{2\pi\iu}\bra[\Psi_\out] (\C_\lambda+\iu \epsilon)^{-1} \ket[\Psi_\inn], \quad \frac{1}{2\pi\iu} \physm{\Psi_\out}{\Psi_\inn}:= \bra[\Psi_\out] (\C_\lambda-\iu \epsilon)^{-1} \ket[\Psi_\inn].
$$
Our next step is to write this expression as a perturbative series in $\lambda$. We will assume that
\be\label{eq:condition}
\lambda \norm{\CE(\CL\pm\iu \epsilon)^{-1}}<1
\ee
for all sufficiently small $\epsilon$. This allows us to express $\physp{\Psi_\out}{\Psi_\inn}$ and $\physm{\Psi_\out}{\Psi_\inn}$ using von-Neumann series (see for example Theorem 3.29 in \cite{Derezinski}):
\be\label{eq:physpm}
\physpm{\Psi_\out}{\Psi_\inn}=\frac{1}{2\pi\iu} \sum_{M=0}^{\infty} \lambda^M (-1)^{M} \bra[\Psi_\out] (\CL\pm\iu \epsilon)^{-1} (\CE (\CL\pm\iu\epsilon)^{-1})^{M} \ket[\Psi_\inn].
\ee
After inserting decompositions of identity before and after each appearance of the operator $\CE$ we obtain the formula (3.20) from \cite{Localsfexpansion} when $\lambda=1$. Let us note, that we assumed that the operators $\C,\CE,\CL$ are bounded and satisfy \eqref{eq:condition}, whereas in \cite{Localsfexpansion} such conditions where not specified.  On the other hand the derivation in \cite{Localsfexpansion} as well as in this paper are formal. We leave the problem of the convergence of the series in specific physical problems for further research.
\subsection{Gravity coupled to massless scalar field}\label{sc:derivscfield}
\renewcommand{\CE}[1][ ]{\hat{K}_{#1}}
\renewcommand{\CL}[1][ ]{\hat{D}_{#1}}
For gravity coupled to a scalar field the scalar constraint is a sum of the matter Hamiltonian and gravitational constraint:
$$
\C_\lambda=\p^2-\CL -\lambda \CE,
$$
where again $\CL$ is the diagonal part of the $\Cgr_\lambda$ and $\lambda\CE$ is its off-diagonal part. We follow the operator approach from the previous section and derive the spin-foam representation of the physical scalar product, where the free part is now $\hat{H}_0=\p^2-\CL$ and the perturbation is $\hat{V}=-\CE$:
\begin{eqnarray}
\phys{\Psi_\out}{\Psi_\inn}:=\frac{1}{2\pi \iu}\lim_{\epsilon\to 0}\left( \bra[\Psi_\out] (\C_\lambda-\iu \epsilon)^{-1} \ket[\Psi_\inn]-\bra[\Psi_\out] (\C_\lambda+\iu \epsilon)^{-1} \ket[\Psi_\inn]\right)=&\\=\lim_{\epsilon\to 0}\left(\physm{\Psi_\out}{\Psi_\inn}-\physp{\Psi_\out}{\Psi_\inn}\right).& \label{eq:physphyspm}
\end{eqnarray}
Using von-Neumann series we expand this expression in powers of the parameter $\lambda$:
\be\label{eq:scphyspm}
\physpm{\Psi_\out}{\Psi_\inn}=\frac{1}{2\pi\iu} \sum_{M=0}^{\infty} \lambda^M \bra[\Psi_\out] (\p^2-\CL\pm\iu \epsilon)^{-1} (\CE (\p^2-\CL\pm\iu\epsilon)^{-1})^{M} \ket[\Psi_\inn].
\ee
As in \cite{AshtekarSpinFoamII} we use an eigenbasis $\ket[\nu,\phi]$ of the volume operator and the scalar field operator. In order to rederive the results from \cite{AshtekarSpinFoamII} we will consider the physical scalar product
\be\label{eq:scphysmat}
\phys{\nu_\out,\phi_\out}{\nu_\inn,\phi_\inn}=\bra[\nu_\out,\phi_\out] 2\p\, \theta(\p) \delta(\C) \ket[\nu_\inn,\phi_\inn],
\ee
where $\theta$ is the Heaviside distribution. We focus first on 
\begin{eqnarray}
\physpm{\nu_\out,\phi_\out}{\nu_\inn,\phi_\inn}=\nonumber \\=\frac{1}{2\pi\iu} \sum_{M=0}^{\infty} \lambda^M \bra[\nu_\out,\phi_\out]2\p\, \theta(\p) (\p^2-\CL\pm\iu \epsilon)^{-1} (\CE (\p^2-\CL\pm\iu\epsilon)^{-1})^{M} \ket[\nu_\inn,\phi_\inn].\label{eq:expansion}
\end{eqnarray}
After inserting the decomposition of identity $\I=d\ep \ket[\ep]\bra[\ep]$ in \eqref{eq:expansion} and taking into account that $\braket{\ep}{\phi}=e^{-\iu \ep \phi}$ we obtain
\be
\label{eq:sfexpansionpI}
\physpm{\nu_\out,\phi_\out}{\nu_\inn,\phi_\inn}=\sum_{M=0}^{\infty} \lambda^M \sum_{{\nu_{M-1},\ldots, \nu_1 \atop \nu_m\neq \nu_{m+1}}} A_{\pm,\epsilon}(\nu_M,\ldots,\nu_0;\phi_f,\phi_i),
\ee
where 
\be
\label{eq:sfexpansionpII}
A_{\pm,\epsilon}(\nu_M,\ldots,\nu_0;\phi_f,\phi_i)=\frac{1}{2\pi \iu}\int_{-\infty}^{\infty} d\ep\, 2\ep\, \theta(\ep) \frac{\eCE[\nu_M \nu_{M-1}]\ldots \eCE[\nu_1 \nu_{0}]}{\prod_{m=0}^{M}(\ep^2-\eCL[\nu_m]\pm \iu \epsilon)} e^{\iu \ep(\phi_\out-\phi_\inn)}.
\ee
This integral can be evaluated using the contour method and the limit of $\epsilon$ going to $0^+$ can be calculated (we assume that the limit and the sum over $M$ can be interchanged). As a result we obtain:
\begin{equation}\label{eq:integrated_expresion}
\fl \phys{\nu_\out,\phi_\out}{\nu_\inn,\phi_\inn}=\sum_{M=0}^{\infty} \lambda^M \sum_{{\nu_{M-1},\ldots, \nu_1 \atop \nu_m\neq \nu_{m+1}}} \eCE[\nu_M \nu_{M-1}]\ldots \eCE[\nu_1 \nu_{0}]\sum_{k=1}^p \frac{1}{(n_k-1)!}\frac{d^{n_k-1}}{d\eCL[w_k]^{n_k-1}}\frac{e^{\iu \sqrt{\eCL[w_k]}(\phi_f-\phi_i)}}{\prod_{m\neq k}^p(\eCL[w_k]-\eCL[w_{m}])^{n_{m}}}.
\end{equation}
The details of the calculation are in \ref{sc:Appendix_A}.

This formula is structurally very similar to the formula (3.27) from \cite{AshtekarSpinFoamII}:
\begin{eqnarray*}
\fl\phys{\nu_\out,\phi_\out}{\nu_\inn,\phi_\inn}=\\\fl=\sum_{M=0}^{\infty} \lambda^M  \sum_{{\nu_{M-1},\ldots, \nu_1 \atop \nu_m\neq \nu_{m+1}}} \eCE[\nu_M \nu_{M-1}]\ldots \eCE[\nu_1 \nu_{0}] \prod_{m=1}^p \frac{1}{(n_m-1)!}\left(\frac{\partial}{\partial \eCL[w_m]}\right)^{n_m-1} \sum_{k=1}^p \frac{e^{\iu \sqrt{\eCL[w_k]}(\phi_\out-\phi_\inn)}}{\prod_{j\neq k}^p(\eCL[w_k]-\eCL[w_{j}])}.
\end{eqnarray*}
In fact they coincide, because:
\begin{eqnarray*}
\prod_{m=1}^p \frac{1}{(n_m-1)!}\left(\frac{\partial}{\partial \eCL[w_m]}\right)^{n_m-1} \sum_{k=1}^p \frac{e^{\iu \sqrt{\eCL[w_k]}(\phi_\out-\phi_\inn)}}{\prod_{j\neq k}^p(\eCL[w_k]-\eCL[w_{j}])}=\\=\sum_{k=1}^p \prod_{m=1}^p \frac{1}{(n_m-1)!}\left(\frac{\partial}{\partial \eCL[w_m]}\right)^{n_m-1} \frac{e^{\iu \sqrt{\eCL[w_k]}(\phi_\out-\phi_\inn)}}{\prod_{j\neq k}^p(\eCL[w_k]-\eCL[w_{j}])}=\\=\sum_{k=1}^p  \frac{1}{(n_k-1)!} \left(\frac{\partial}{\partial \eCL[w_k]}\right)^{n_k-1} \left(e^{\iu \sqrt{\eCL[w_k]}(\phi_\out-\phi_\inn)}\prod_{j\neq k}^p { \frac{1}{(n_j-1)!}\left(\frac{\partial}{\partial \eCL[w_j]}\right)^{n_j-1}} \frac{1}{\eCL[w_k]-\eCL[w_{j}]} \right)=\\=\sum_{k=1}^p  \frac{1}{(n_k-1)!} \left(\frac{\partial}{\partial \eCL[w_k]}\right)^{n_k-1} \frac{e^{\iu \sqrt{\eCL[w_k]}(\phi_\out-\phi_\inn)}}{\prod_{j\neq k}^p (\eCL[w_k]-\eCL[w_{j}])^{n_j}}.
\end{eqnarray*}

Let us notice that the scalar product used in \cite{AshtekarSpinFoamII} was
$$
\phys{\nu_\out,\phi_\out}{\nu_\inn,\phi_\inn}=\bra[\nu_\out,\phi_\out] 2|\p|\, \delta(\C) \ket[\nu_\inn,\phi_\inn].
$$ 
Since $|\p|=\p\theta(\p)-\p\theta (-\p)$, it is straightforward to apply our result and recover their formula (3.26). However, the authors next restrict to positive frequency part and arrive at their final formula (3.27). Our calculation shows explicitly that this is equivalent to considering the scalar product
$$
\phys{\nu_\out,\phi_\out}{\nu_\inn,\phi_\inn}=\bra[\nu_\out,\phi_\out] 2\p \theta(\p)\, \delta(\C) \ket[\nu_\inn,\phi_\inn].
$$
\subsubsection{Functional calculus}\label{sc:derivcan}
We will present now very short derivation of the formula \eqref{eq:integrated_expresion} using the functional calculus. On the one hand the expansion physical scalar product \eqref{eq:scphysmat} studied in the previous subsection has covariant interpretation. On the other hand, it can be written as matrix elements of an evolution operator:
\begin{equation}\label{eq:physcosmo}
\phys{\nu_\out,\phi_\out}{\nu_\inn,\phi_\inn}=\bra[\nu_\out] e^{\iu \sqrt{\Cgr_\lambda} (\phi_\out-\phi_\inn) } \ket[\nu_\inn].
\end{equation}
By the functional calculus (see for example Definition 1.10.1 of \cite{DerezinskiU}):
$$
\phys{\nu_\out,\phi_\out}{\nu_\inn,\phi_\inn}=\frac{1}{2\pi \iu}\int_{\gamma} dE \bra[\nu_\out] (E-\Cgr_\lambda)^{-1} e^{\iu \sqrt{E} (\phi_\out-\phi_\inn) } \ket[\nu_\inn],
$$
where the contour $\gamma$ encircles the spectrum of $\C$ counterclockwise. Let us notice that this integral coincides with the integral over $\ep$ under the substitution $E=\ep^2$: the integral in $A_{+,\epsilon}(\nu_M,\ldots,\nu_0;\phi_\out,\phi_\inn)$ is the part of the contour above the spectrum and $A_{-,\epsilon}(\nu_M,\ldots,\nu_0;\phi_\out,\phi_\inn)$ is the part below the spectrum. Again using von-Neumann series we arrive at:
\begin{eqnarray*}
\fl\phys{\nu_\out,\phi_\out}{\nu_\inn,\phi_\inn}=\frac{1}{2\pi \iu}\int_{\gamma} dE \bra[\nu_\out]\sum_{M=0}^{\infty} \lambda^M (E-\CL)^{-1} (\CE (E-\CL)^{-1})^M e^{\iu \sqrt{E} (\phi_\out-\phi_\inn) } \ket[\nu_\inn]=\\= \sum_{M=0}^{\infty} \lambda^M  \sum_{{\nu_{M-1},\ldots, \nu_1 \atop \nu_m\neq \nu_{m+1}}} A(\nu_M,\ldots,\nu_0; \phi_\out, \phi_\inn),
\end{eqnarray*}
where
\begin{eqnarray}
A(\nu_M,\ldots,\nu_0; \phi_\out, \phi_\inn)=\frac{1}{2\pi \iu}  \eCE[\nu_M \nu_{M-1}]\ldots \eCE[\nu_1 \nu_{0}]  \int_\gamma dE \frac{e^{\iu \sqrt{E} (\phi_\out-\phi_\inn)}}{\prod_{m=0}^M (E-\eCL[\nu_m])}=\\=\frac{1}{2\pi \iu}  \eCE[\nu_M \nu_{M-1}]\ldots \eCE[\nu_1 \nu_{0}]  \int_\gamma dE \frac{e^{\iu \sqrt{E} (\phi_\out-\phi_\inn)}}{\prod_{m=0}^p (E-\eCL[w_m])^{n_m}}=\\=\eCE[\nu_M \nu_{M-1}]\ldots \eCE[\nu_1 \nu_{0}]\sum_{k=1}^p\Res(\frac{e^{\iu \sqrt{E} (\phi_\out-\phi_\inn)}}{\prod_{m=0}^p (E-\eCL[w_m])^{n_m}},\eCL[w_k])=\label{eq:sfexpansionE}\\= \eCE[\nu_M \nu_{M-1}]\ldots \eCE[\nu_1 \nu_{0}] \sum_{k=1}^p \frac{1}{(n_k-1)!}\frac{d^{n_k-1}}{d\eCL[w_k]^{n_k-1}}\frac{e^{\iu \sqrt{\eCL[w_k]}(\phi_\out-\phi_\inn)}}{\prod_{m\neq k}^p(\eCL[w_k]-\eCL[w_{m}])^{n_{m}}}.
\end{eqnarray}
Since the theory of perturbations of the spectra is well studied mathematically (see for example \cite{ReedSimonIV}), we expect that the functional calculus approach presented in this subsection may be useful to study the convergence of the series analytically. We leave such study for further research and focus on formal derivation of the spin foam models from the canonical theory.
\renewcommand{\CE}[1][ ]{\hat{C}_{{\rm E }#1}}
\renewcommand{\eCE}[1][ ]{C_{{\rm E }#1}}
\renewcommand{\CL}[1][ ]{\hat{C}_{{\rm L }#1}}
\renewcommand{\eCL}[1][ ]{C_{{\rm L }#1}}
\newcommand{\sCE}[1][ ]{\hat{C}^*_{{\rm E }#1}}
\section{Spin-foam paradigm in Loop Quantum Gravity}\label{sc:LQG}
\subsection{The scalar constraint operator}\label{sc:scalar_constraint}
The Hamiltonian analysis of General Relativity leads to a system with vanishing true Hamiltonian and three constraints: Gauss constraint, vector constraint and scalar constraint. The Gauss constraint generates SU(2) gauge transformations. The Hilbert space of solutions to the Gauss constraint is spanned by SU(2)-invariant states called spin-network states (see for example \cite{StatusReport,ThiemannBook, BaezSN}):
$$
\ket[\gamma,\rho,\iota],
$$
where $\gamma$ is an oriented graph, $\rho$ is a coloring of the links of the graph with unitary irreducible representations of the SU(2) group and $\iota$ is a coloring of the nodes of the graph with tensors invariant under the action of the group:
$$
\iota_n \in \Inv{\Hil_{\rho_{\link_1}^*}\otimes\ldots\otimes \Hil_{\rho_{\link_M}^*}\otimes \Hil_{\rho_{\link_{M+1}}}\otimes\ldots \otimes \Hil_{\rho_{\link_{N}}}},
$$
where links $\link_1,\ldots, \link_M$ are incoming to the node $n$ and links $\link_{M+1},\ldots, \link_N$ are outgoing from the node $n$.

We will say that two spin networks $s=(\gamma,\rho,\iota)$ and $s'=(\gamma',\rho',\iota')$ are equivalent if there is a spin network $s''=(\gamma'',\rho'',\iota'')$ that can be obtained from $s$ and $s'$ by sequences of operations of flipping orientation links, splitting links, adding links and adding nodes (see \cite{BaezSN,SFLQG, KisielowskiPhD}). We will write $s\sim s'$ and $\gamma\sim \gamma'$. By the same symbol we will denote an equivalence of group representations. In particular, $\rho_{\link}\sim \rho'_{\link}$ will mean that there is an linear isomorphism $\omicron_\link:\Hil_{\rho_{\link}}\to \Hil_{\rho'_{\link}}$ such that $\rho'_{\link}(g)\circ \omicron_\link= \omicron_\link\circ \rho_{\link}(g)$ for each $g\in{\rm SU(2)}$.

A scalar product between two spin-network states $s_1$ and $s_2$ is non-zero only if there are spin-networks $s'_1$ and $s'_2$ equivalent to $s_1$ and $s_2$, respectively, defined on the same graph $\gamma$. Since the operations on the spin networks preserve the scalar product, the remaining property defining the scalar product is:
\begin{equation}\label{eq:snscalarproduct}
\braket{\gamma,\rho',\iota'}{\gamma,\rho,\iota}= \delta_{\rho,\rho'}\prod_{\link\in\links{\gamma}} \frac{1}{\dim \rho_\link}\prod_{\node\in \nodes{\gamma}}\braket{\iota'_n}{\omicron_\node\iota_n},
\end{equation}
where 
$$
\delta_{\rho,\rho'}=\cases{1,&if $\forall_{\link\in \links{\gamma}} \rho_\link \sim \rho'_\link$,\\
0&otherwise,\\}
$$
and 
\begin{equation}\label{eq:omicron}
\omicron_\node=(\omicron_{\link_1}^{-1})^*\otimes \ldots \otimes (\omicron_{\link_M}^{-1})^* \otimes \omicron_{\link_{M+1}}\otimes \ldots \otimes \omicron_{\link_{N}}.
\end{equation}

Following \cite{LewandowskiSahlmann} let us denote by $\tilde{\Hil}_{\gamma}$ the subspace spanned by all spin-network states defined on a graph $\gamma$. Let $\gamma$ be a graph obtained from $\gamma'$ by the operations discussed above. Due to the equivalence of spin networks $\tilde{\Hil}_{\gamma}$ is a proper subspace of $\Hil_{\gamma'}$:
$$
\tilde{\Hil}_{\gamma'}< \tilde{\Hil}_{\gamma}.
$$
We will say that a spin network state $\ket[s] \in \tilde{\Hil}_{\gamma}$ is proper iff:
$$
\tilde{\Hil}_{\gamma'}< \tilde{\Hil}_{\gamma} \Rightarrow \ket[s] \perp \tilde{\Hil}_{\gamma'}.
$$
The space spanned by proper spin networks defined on the graph $\gamma$ will be denoted by $\Hil_\gamma$.

Following \cite{DziendzikowskiDomagalaLewandowski} and \cite{LewandowskiSahlmann} we consider a Hilbert space of solutions to the Gauss constraint and partial solutions to the vector constraint $\Hvgr$. An action of a diffeomorphism $f:\slice\to\slice$ on a spin-network state $\ket[\gamma,\rho,\iota]$ is given by: 
$$
U_f \ket[\gamma,\rho,\iota] = \ket[f(\gamma),\rho',\iota'],
$$
where $\rho'_{f(\link)}=\rho_{\link}, \iota'_{f(\node)}=\iota_{\node}$. Let us denote by $\TDiff{\gamma}$ the set of diffeomorphism that act trivially on $\Hil_{\gamma}$ and by $\DV$ the set of diffeomorphisms that act trivially on a finite subset $V=\{x_1,\ldots, x_n\}$ of the space manifold $\slice$. A basis of the space $\Hvgr$ is formed by states
$$
\ket[{[\gamma,\rho,\iota]}]
$$
obtained from $\ket[\gamma,\rho,\iota]$ by averaging over all diffeomorphisms $\Diff_{\nodes{\gamma}}$ modulo $\TDiff{\gamma}$. The scalar product of two such states is given by
\begin{equation}\label{eq:partialdifffix}
\braket{{[\gamma',\rho',\iota']}}{{[\gamma,\rho,\iota]}}:=\frac{1}{N_\gamma}\sum_{[f]\in \Diff_{\nodes{\gamma}}\slash \TDiff{\gamma}: \gamma'=f(\gamma)} \bra[\gamma',\rho',\iota'] U_f \ket[\gamma,\rho,\iota],
\end{equation}
where $N_\gamma$ is a free constant which we fix to be equal to $1$ \footnote{Our convention differs from the convention from \cite{LewandowskiSahlmann}, where $N_\gamma$the number of equivalence classes $[f]\in\Diff_{\nodes{\gamma}}\slash \TDiff{\gamma}$ such that $f(\gamma)=\gamma$.}. The space $\Hvgr$ can be decomposed into a direct sum of Hilbert spaces $\HVgr$, where $V=\{x_1,\ldots, x_n\}$ ranges over all finite subsets of the space manifold $\slice$. In this space the scalar constraint for the gravitational field $\Cgr_x$ can be defined \cite{LewandowskiSahlmann,NewScalarConstraint}. It consists of an Euclidean $\CE$ and Lorentzian $\CL$ part:
$$
\Cgr_x=\CL[x] + \lambda \CE[x],
$$
where the value of $\lambda$ is determined by the Barbero-Immirzi parameter as in \cite{TimeEvolution}. In this paper we will focus on \cite{NewScalarConstraint} because in this proposal the Lorentzian part is graph-preserving. In order to pass to this approach we will recall the basic properties of the operators $\CL[x_I]$ and $\CE[x_I]$, where we use the version from \cite{NewScalarConstraint}. The operators $\CL[x_I]$ are first defined on the Hilbert space of SU(2) gauge-invariant cylindrical functions, which is spanned by the spin-network states. When acting on a spin-network state, the operator does not change its graph nor the labels of the links with irreducible representations. 

Let $\R=(\rho_1,\ldots, \rho_N)$ be a sequence of representations of the SU(2) group. Given a Hilbert space of invariant tensors 
$$
\Hil_\R=\Inv{ \Hil_{\rho_1}\otimes \ldots \otimes \Hil_{\rho_N}}
$$
we define operators $\J_{r\, i},\, r\in \{1,\ldots, N\}, i\in\{1,2,3\}$ by the following formula:
$$
\J_{r\, i} := \I\otimes \ldots\otimes \I\otimes \rho'_r(\tau_i)\otimes \I \otimes \ldots \otimes \I,
$$
where $\tau_i=-\frac{\iu}{2}\sigma_i$ is the basis of the su(2) Lie algebra defined by the Pauli matrices $\sigma_i$ and $\rho'_r$ is the representation of su(2) corresponding to $\rho_r$. For each pair $(\R,\varepsilon)$ of a sequence $\R$ and a symmetric function $\epsilon:\{1,\ldots,N\}\times\{1,\ldots,N\} \to \{0,1\}$ we define an operator $\CL[(\R,\varepsilon)]:\Hil_\R\to \Hil_\R$ by the following formula:
$$
\CL[(\R,\varepsilon)]=\sum_{r,s} \varepsilon_{rs}\sqrt{\sum_{i=1}^3(\epsilon_{ijk} \J_{r\, j} \J_{s\, k} )^2} \left(\pi + \arccos \left( \frac{\sum_{i=1}^{3}\J_{r\, i}\J_{s\, i}}{\sqrt{\sum_{i=1}^{3}\J_{r\, i}^2}\sqrt{\sum_{i=1}^{3}\J_{s\, i}^2}} \right) \right).
$$
Let us note that the order of representations $\R$ is not important. Given a spin network $s=(\gamma,\rho,\iota)$, we associate with each node $x_I$ a sequence of of links representations $(\link_1,\ldots,\link_M,\link_{M+1},\ldots,\link_N)$, where $\link_1,\ldots,\link_M$ are incoming to the node $x_I$ and $\link_{M+1},\ldots,\link_N$ are outgoing from the node $x_I$, and a sequence of representations
$$\R_{x_I}=(\rho^*_{\link_1},\ldots,\rho^*_{\link_{M}},\rho_{\link_{M+1}},\ldots,\rho_{\link_N}),$$
 Moreover, for each node $x_I$ we define a function assigning each pair $(r,s)\in \{1,\ldots,N\}\times \{1,\ldots,N\}$ a number 0 or 1 according to the following rule:
\begin{equation}\label{eq:epsilonrs}
\varepsilon_{x_I,\,rs}=\cases{1,&if the tangent vectors to the links $\link_r$ and $\link_s$ at $\node$\\ & are linearly independent,\\
0&otherwise.\\}
\end{equation}
Let $\CL[x_I]$ be an operator acting on the space of SU(2) invariant tensors associated to the node $x_I$ that is defined by
$$
\CL[x_I]:=\CL[(\R_{x_I},\varepsilon_{x_I})].
$$
The action of the Lorentzian part of the scalar constraint on a state $\ket[{[\gamma,\rho,\iota]}]$ is given by
$$
\CL[x_I] \ket[{[\gamma,\rho,\iota]}]=\ket[{[\gamma,\rho,{\CL[x_I]}\iota]}],
$$
where
$$
(\CL[x_I]\iota)_\node=\cases{\CL[\node]\iota_\node,& if $x_I=\node$,\\ \iota_n & otherwise.\\}
$$
This defines the action of $\CL[x_I]$ for states such that $x_I \in \nodes{\gamma}$. By cylindrical equivalence of spin networks, it can be defined for arbitrary spin networks -- this amounts to extending the operator defined above by $0$ to all states.

\begin{figure}
\begin{center}
\includegraphics[scale=0.7]{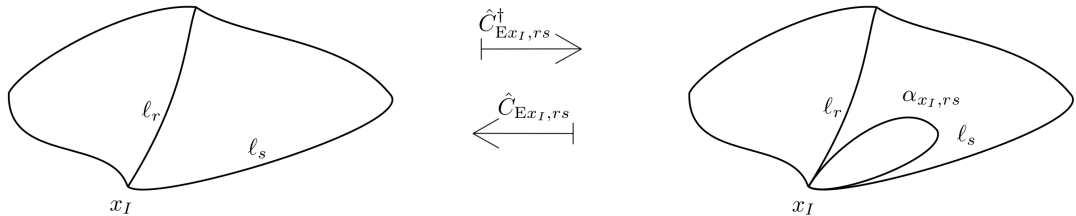}
\caption{The operator $\CE[x_I, rs]^{\dagger}$ adds a loop $\alpha_{x_I,rs}$ tangent to the links $\link_r$ and $\link_s$ at $x_I$ and the operator $\CE[x_I, rs]$ removes such loop.}
  \label{fig:CE}\end{center}
\end{figure}

The Euclidean part $\CE$ is graph-changing. According to the prescription from \cite{HamiltonianOperator,NewScalarConstraint} it adds and subtracts loops tangential to links of the graph. Let us describe the action of this operator on a state $\ket[{\gamma,\rho,\iota}]$. Let us denote by $\CE[x_I, rs]^\dagger\gamma$ the graph obtained from $\gamma$ by adding a loop $\alpha_{x_I, rs}$ tangential to the links $\link_r$ and $\link_s$ at the node $x_I$ oriented such that its beginning is tangent to the link $\link_r$ and its end is tangent to the link $\link_s$ (see figure \ref{fig:CE}). By $\CE[x_I, rs]\gamma$ we denote the graph obtained from $\gamma$ by removing such loop (if possible) \footnote{In the formulation we will present below it is not relevant whether $\CE[x_I, rs]^\dagger$ adds or subtracts a loop but we choose a convention compatible with \cite{HamiltonianOperator,NewScalarConstraint}}. Let us denote by $\CE[x_I, rs]^\dagger \rho$ the following coloring of the links of the graph $\CE[x_I, rs]^\dagger \gamma$:
$$
(\CE[x_I, rs]^\dagger \rho)_{\link} = \cases{\rho_{\link} & if $\link$ belongs to the graph $\gamma$,\\ \rho_{(l)}& otherwise.}
$$
By $\CE[x_I, rs] \rho$ we denote the following coloring of the graph $\CE[x_I, rs]\gamma$:
$$
(\CE[x_I, rs]\rho)_{\link}=\rho_{\link}.
$$

Let $\R=(\rho_1,\ldots, \rho_N)$ be a sequence of representations of the SU(2) group. Denote by $\R_{(l)}$ the sequence
$$
\R_{(l)}=(\rho_{(l)}, \rho_{(l)}^{*},\rho_1,\ldots, \rho_N).
$$
Let $\J^i_{(l)}={\rho'}_{(l)}(\tau_i):\Hil_{(l)}\to \Hil_{(l)}$. Clearly $\J^i_{(l)}$ can be considered to be an element of $\Hil_{(l)}\otimes \Hil_{(l)}^*$. We define an operator $\CE[(R,rs)]^\dagger:\Hil_\R \to \Hil_{\R_{(l)}}$ by the following formula:
\begin{equation}
\CE[(\R,rs)]^\dagger=-\frac{3}{l(l+1)(2l+1)}\epsilon_{ijk} \J^i_{(l)} \J_{r}^j \J_{s}^k.
\end{equation}
Let us explain in more detail how the operator $\CE[(\R,rs)]^\dagger$ acts on a state $\iota\in\Hil_\R$ by using the abstract index notation:
$$
(\CE[(\R,rs)]^\dagger \iota)^{C_1, B_1\ldots B_N}_{C_2} = {\CE[(\R,rs)]^\dagger}^{C_1, B_1\ldots B_N}_{C_2, A_1\ldots A_N} \iota^{A_1\ldots A_N},
$$
where the indices $C_1$ and $C_2$ correspond to the spaces $\Hil_{(l)}$ and $\Hil_{(l)}^*$, respectively; $A_1,\ldots, A_N$ and $B_1,\ldots, B_N$ are indices corresponding to $\Hil_\R$. Let us denote by $\CE[(\R,rs)]:\Hil_{\R_{(l)}} \to \Hil_\R$ the operator adjoint to $\CE[(\R,rs)]^\dagger$. 
 
In the following we will make a simplifying assumption. Let us denote by $\mathring{\gamma}$ a graph obtained from $\gamma$ by removing all loops tangential to two non-tangential links of $\gamma$. We limit to spin networks $s=(\gamma,\rho,\iota)$ such that the only diffeomorphism satisfying
\begin{equation}\label{eq:simplification}
f(\mathring{\gamma})=\mathring{\gamma},\quad \forall_{\link\in\links{\mathring{\gamma}}}\rho_{f(\link)}=\rho_{\link}
\end{equation}
is the identity diffeomorphism. This restriction is not present in other papers on this subject 
(see for example \cite{GravityQuantized, DziendzikowskiDomagalaLewandowski, HamiltonianOperator, LewandowskiSahlmann}). It is a simplifying assumption that makes our presentation clearer. For example, with this assumption the averaged scalar product $\braket{[\gamma,\rho',\iota']}{[\gamma,\rho,\iota]}$ reduces to the standard scalar product $\braket{\gamma,\rho',\iota'}{\gamma,\rho,\iota}$:
$$
\braket{[\gamma,\rho',\iota']}{[\gamma,\rho,\iota]}=\braket{\gamma,\rho',\iota'}{\gamma,\rho,\iota}.
$$

We define an operator $\CE[x_I, rs]^\dagger$ by its action on a state $\ket[{[\gamma,\rho,\iota]}]$:
$$
\CE[x_I,rs]^\dagger \ket[{[\gamma,\rho,\iota]}]=\ket[{\left[\CE[(\R_{x_I},rs)]^\dagger\gamma,\CE[(\R_{x_I,{rs}})]^\dagger \rho, \CE[(\R_{x_I,{rs}})]^\dagger \iota \right]}].
$$
Let us note that the operator $\CE[x_I,rs]^\dagger$ can be defined on states averaged over diffeomorphisms that do not move the nodes of the graph due to our simplifying assumption \eqref{eq:simplification}. With this assumption links $r$ and $s$ in the averaged state are distinguishable by the labeling with representations. The adjoint operator is:
\begin{equation}\label{eq:CEadj}
\CE[x_I,rs] \ket[{[\gamma,\rho,\iota]}]=\frac{1}{2l+1}\ket[{\left[\CE[(\R_{x_I},rs)]\gamma,\CE[(\R_{x_I,{rs}})] \rho, \CE[(\R_{f(x_I)},rs)] \iota \right]}],
\end{equation}
where 
$$
\ket[{\left[\CE[(\R_{x_I},rs)]\gamma,\CE[(\R_{x_I,{rs}})]\rho, \CE[(\R_{x_I,{rs}})] \iota \right]}] :=0
$$
if $\gamma$ has no loops tangent to links $\link_r$ and $\link_s$ at the node $x_I$. The factor $\frac{1}{2l+1}$ comes from the fact that the scalar product in the spin-network basis is given by Haar integrals over the group normalized to $1$, which will be shown in equation \eqref{eq:Cdagger}. Let $\gamma'$ be a graph obtained from $\gamma$ by adding a loop tangential to the links $\link_r$ and $\link_s$ at the node $x_I$ oriented such that its beginning is tangent to the link $\link_r$ and its end is tangent to the link $\link_s$, let $\rho'$ be a coloring of the links of the graph $\gamma'$ such that $\rho'_\link =\rho_\link$. 
\begin{eqnarray}\label{eq:Cdagger}
 \brakett{{[\gamma',\rho',\iota']}\left| \CE[(x_I, rs)]^\dagger{ [\gamma,\rho,\iota]}\right.}=\\=\sum_{[f]\in \Diff_{\nodes{\CE[(x_I, rs)]^\dagger\gamma}}\slash \TDiff{\CE[(x_I, rs)]^\dagger\gamma}: \gamma'=f(\CE[(x_I, rs)]^\dagger\gamma)}\braket{\iota'_{x_I}}{ \CE[(\R_{x_I},rs)]^\dagger \iota_{x_I}} \frac{1}{\dim \rho_{\alpha_{rs}}}\\ \prod_{\link\in \links{\gamma}}\frac{\delta_{\rho'_{f(\link)},\rho_{\link}}}{\dim \rho_{\link}}  \prod_{\node\in\nodes{\gamma},\node\neq x_I} \braket{\iota'_{n}}{ \iota_{n}}=\\=\frac{1}{2l+1} \sum_{[f']\in \Diff_{\nodes{\gamma}}\slash \TDiff{\gamma}: \CE[(x_I, rs)] \gamma' = f'(\gamma)} \braket{\CE[(\R_{x_I},rs)]\iota'_{x_I}}{ \iota_{x_I}} \\ \prod_{\link\in \links{\gamma}}\frac{\delta_{\rho'_{f'(\link)},\rho_{\link}}}{\dim \rho_{\link}}  \prod_{\node\in\nodes{\gamma},\node\neq x_I} \braket{\iota'_{n}}{\iota_{n}}= \brakett{{\CE[(x_I, rs)] [\gamma',\rho',\iota']}\left|{ [\gamma,\rho,\iota]}\right.}
\end{eqnarray}
Thanks to the assumptions \eqref{eq:simplification} on the spin networks (and about the graphs $\gamma$ and $\gamma'$ made above) the sets
$$\{[f]\in \Diff_{\nodes{\CE[(x_I, rs)]^\dagger\gamma}}\slash \TDiff{\CE[(x_I, rs)]^\dagger\gamma}: \gamma'= f(\CE[(x_I, rs)]^\dagger\gamma), \forall_{\link\in \links{\gamma}}\rho'_{f(\link)}=\rho_\link\}$$ 
and
$$ \{[f']\in \Diff_{\nodes{\gamma}}\slash \TDiff{\gamma}: \CE[(x_I, rs)] \gamma'= f'(\gamma), \forall_{\link\in\links{\gamma}} \rho'_{f'(\link)}=\rho_\link \}$$
contain precisely 1 element and therefore the sums over the diffeomorphisms in \eqref{eq:CEadj} have only 1 non-trivial term equal for both sums.

The operator $\CE[x_I]$ is defined by
$$
 \CE[x_I]\ket[{[\gamma,\rho,\iota]}]:= \sum_{r,s} \varepsilon_{x_I, rs}(\CE[x_I, rs]+\CE[x_I, rs]^\dagger)\ket[{[\gamma,\rho,\iota]}],
$$
where $\varepsilon_{x_I,rs}$ is defined in \eqref{eq:epsilonrs}. With this definition it is symmetric (see also \cite{LewandowskiSahlmann}).

\subsection{The physical scalar product}

Suppose that the constraint operators $\Cgr_x$ are essentially self-adjoint. In such case, each of the spaces $\Hil_{\{x_1,\ldots, x_n\}}$ can be decomposed using the spectral decompositions of the operators $\Cgr_{x_I}, I\in\{1,\ldots,n\}$. Let $P^{\rm gr}_{c_{x_I}}$ be the projection-valued measure corresponding to $\Cgr_{x_I}$. Our notation for projection-valued measures is the same as in \cite{ReedSimonI}, in particular:
$$
\C_{x_I}=\int c_{x_I} d P^{\rm gr}_{c_{x_I}}.
$$
We also define:
$$
P^{\rm gr}_{c_{x_1}\ldots c_{x_N}}:=\bigotimes_{I=1}^n P^{\rm gr}_{c_{x_I}}.
$$

For the matter part we consider the polymer quantization. The polymer Hilbert space $\HVmat$ is spanned by functionals of a real-valued scalar field $\varphi:\slice\to \mathbb{R}$ given by:
$$
\ket[\pi][\varphi]=U_{\pi}(\varphi)=e^{\iu (\pi(x_1)\varphi(x_1)+\ldots +\pi(x_n)\varphi(x_n))},
$$ 
where $\pi:\slice\to \mathbb{R}$ is a function with finite support $\supp(\pi)=\{x_1,\ldots,x_n\}$. It is equipped with a scalar product
$$
\braket{U_\pi}{U_{\pi'}}=\delta_{\pi,\pi'},
$$
where $\delta$ is the Kronecker delta. The states are eigenvectors of the momentum operator
$$
\hat{\pi}(V) \ket[\pi] = \left(\sum_{x\in V} \pi(x)\right) \ket[\pi].
$$
Let $P^{\rm mat}_{\pi_{x_I}}$ be the projection operator:
$$
P^{\rm mat}_{\pi_{x_I}}=\ket[\pi_{x_I}]\bra[\pi_{x_I}]
$$
onto the eigenvector $\ket[\pi_{x_I}]$. We also define
$$
P^{\rm mat}_{\pi_{x_1}\ldots \pi_{x_n}}=\bigotimes_{I=1}^n P^{\rm mat}_{\pi_{x_I}}
$$

We consider the Hilbert space 
$$
\HV=\HVmat \otimes \HVgr
$$ 
and a scalar constraint:
$$
\C_{x_I}= \hat{\pi}_{x_I}^2 - \Cgr_{x_I}.
$$
The space of solutions of the constraint is described by the physical scalar product. Let us consider two states $\Psi_\inn\in \HV$ and $\Psi_\out\in \Hil_{V'}$. The physical scalar product is given by:
$$
\phys{\Psi_\out}{\Psi_\inn}=\int \sum_{\pi_{x_1},\ldots,\pi_{x_n}} \prod_{I=1}^n\delta_{\pi_{x_I}, \sqrt{c_{x_I}}} \braket{\tilde{\eta}(\Psi_\out)}{P^{\rm mat}_{\pi_{x_1}\ldots \pi_{x_n}}\otimes d P^{\rm gr}_{c_{x_1}\ldots c_{x_n}} \Psi_\inn},
$$
where $\tilde{\eta}$ corresponds to averaging with respect to the remaining diffeomorphisms ${\Diff\slash \Diff_{V'}}$. By choosing $\delta_{\pi_{x_I}, \sqrt{c_{x_I}}}$ we restricted to positive frequencies (momenta). Our derivation can be easily generalized to include the negative frequencies as well. The integral is a multiple integral over the spectra of $\Cgr_{x_I}$ and the sum ranges over the sets of real numbers, i.e. the spectra of $\hat{\pi}_{x_I}$. The Kronecker deltas $\delta_{\pi_{x_I}, \sqrt{c_{x_I}}}$ impose the scalar constraint. Since the scalar constraint operator does not change the set $V$, the physical scalar product is non-zero only if $V'=f(V)$ for a diffeomorphism $f$. Without loss of generality, we can focus on the case $V=V'$. In this case the equivalence classes of ${\Diff\slash \Diff_{V}}$ that contribute non-trivially to the scalar product are labelled by permutations of the vertices of $V=\{x_1,\ldots,x_n\}$. In the following we will denote by $\sigma\in S_n$ a permutation of the numbers $\{1,\ldots,n\}$ and a representative of an equivalence ${\Diff\slash \Diff_{V}}$ such that
$$
\sigma(x_I)=x_{\sigma(I)}.
$$
The averaging with respect to the diffeomorphisms ${\Diff\slash \Diff_{V}}$ is defined up to a factor depending on $V$. We choose it to be equal to the inverse of the number of permutations of the vertices of $V$. With this choice a diffeomorphism invariant scalar product between two states $\Psi_\in,\Psi_\out\in \HV$ is given by
$$
\braket{\tilde{\eta}(\Psi_\out)}{\Psi_\inn}=\frac{1}{\NV} \sum_{\sigma\in S_n} \braket{ \Psi_\out}{U_\sigma \Psi_\inn}.
$$

We will study the physical scalar product between states $\ket[\Psi_\inn]$ and $\ket[\Psi_\out]$ that are of the following form
$$
\ket[\Psi_{\inn\slash \out}] = \ket[{[s_{\inn\slash \out}]}] \otimes \ket[\Psi^{\rm mat}_{\inn\slash \out}],
$$
where $\ket[\Psi^{\rm mat}_{\inn\slash \out}]\in L^2(\mathbb{R}^n,\prod_{I=1}^n\Bohr[\varphi_{\inn\slash \out}(x_I)])$. For such states the action of the diffeomorphism $U_\sigma$ splits into diffeomorphisms acting on the gravitational part of the Hilbert space and matter part:
$$
U_{\sigma}=U^{\rm gr}_\sigma \otimes U^{\rm mat}_\sigma.
$$
The diffeomorphisms $\sigma$ permutes the vertices of the (averaged) spin-network state $[\gamma,\rho,\iota]$: 
$$
U^{\rm gr}_\sigma \ket[{[\gamma,\rho,\iota]}]=\ket[{[\sigma(\gamma),\rho,\iota']}],
$$
where $\iota'_{x_{\sigma(I)}}=\iota_{x_I}$ as well as the points of the state $\ket[\Psi^{\rm mat}]$:
$$
(U_\sigma^{\rm mat} \Psi^{\rm mat})(\varphi)=\Psi^{\rm mat}(\sigma^*\,\varphi).
$$

We will be interested in the amplitudes $A([s_\out],\varphi_\out;[s_\inn],\varphi_\inn)$ defined by:
\begin{eqnarray*}
\fl \phys{\Psi_\out}{\Psi_\inn}=:\int \prod_{I=1}^{n}\Bohr[\varphi_\out(x_I)] \Bohr[\varphi_\inn(x_I)] \overline{\Psi_\out^{\rm mat}}(\varphi_\out) A([s_\out],\varphi_\out;[s_\inn],\varphi_\inn) \Psi_\inn^{\rm mat}(\varphi_\inn).
\end{eqnarray*}
Our goal will be to write the amplitude $A([s_\out],\varphi_\out(x_I);[s_\inn],\varphi_\inn(x_I))$ as a spin-foam amplitude. Let us notice that
\begin{eqnarray*}
A([s_\out],\varphi_\out(x_I);[s_\inn],\varphi_\inn(x_I))=\\= \frac{1}{\NV} \sum_{\sigma\in S_n} \int \sum_{\pi_{x_1},\ldots,\pi_{x_n}} \prod_{I=1}^n\delta_{\pi_{x_I}, \sqrt{c_{x_I}}} e^{-\iu \pi_{x_I} (\varphi_\out(x_{\sigma(I)})-\varphi_\inn(x_I)) } \braket{[s_\out]}{ U_\sigma^{\rm gr}\, d P^{\rm gr}_{c_{x_1}\ldots c_{x_n}} |[s_\inn]}=\\= \frac{1}{\NV}\sum_{\sigma\in S_n}\int \prod_{I=1}^n e^{-\iu \sqrt{c_{x_I}} (\varphi_\out(x_\sigma(I))-\varphi_i(x_I)) } \braket{[s_\out]}{U_\sigma^{\rm gr}\, d P^{\rm gr}_{c_{x_1}\ldots c_{x_n}} |[s_\inn]}=\\=\frac{1}{\NV}\sum_{\sigma\in S_n}\braket{[s_\out]}{ U^{\rm gr}_\sigma e^{-\iu\sum_{I=1}^n(\varphi_\out(x_{\sigma(I)})-\varphi_\inn(x_I))\sqrt{\Cgr_{x_I}}} |[s_\inn]}.
\end{eqnarray*}
Let us notice that $A([s_\out],\varphi_\out(x_I);[s_\inn],\varphi_\inn(x_I))$ are just matrix elements of an evolution operator. Since in LQC with massless scalar field the physical scalar product can be also expressed as matrix elements of an evolution operator \eqref{eq:physcosmo}, we can use the formulas \eqref{eq:physphyspm}, \eqref{eq:scphyspm}, \eqref{eq:scphysmat}, \eqref{eq:expansion}, \eqref{eq:sfexpansionpI}, \eqref{eq:sfexpansionpII} to write $A([s_\out],\varphi_\out(x_I);[s_\inn],\varphi_\inn(x_I))$ in the following form:
\begin{eqnarray}
 \fl A([s_\out],\varphi_\out(x_I);[s_\inn],\varphi_\inn(x_I)) =\lim_{\epsilon\to 0^+}\sum_{M=0}^{\infty} \lambda^M \sum_{{M_1,\ldots, M_n\atop M_1+\ldots+M_n=M}}\sum_{\sign} \frac{1}{\NV}\sum_{\sigma \in S_n}  \frac{1}{(\pi\iu)^n}\prod_{I=1}^n\int_{-\infty}^{+\infty}  dp_{x_{I}} p_{x_{I}} \theta(p_{x_{I}}) \\ \fl e^{-\iu p_{x_{I}} (\varphi_\out(x_{\sigma(I)})-\varphi_\inn(x_I))}\brakett{[s_\out] \left| U_{\sigma}^{\rm gr}\prod_{J=1}^n \sign_{x_J} \left(p^2_{x_J}-\CL[x_J]+\iu \epsilon\sign_{x_J}\right)^{-1} \left(\CE[x_J] (p^2_{x_J}-\CL[x_J]+\iu\epsilon\sign_{x_J})^{-1}\right)^{M_J}\right|[s_\inn]},\label{eq:sfoperator}
\end{eqnarray}
where the third sum is over all functions $\sign:V\to \{-1,1\}$. The integrals with respect to $p_{x_I}$ are with the standard Lebesgue measure on $\mathbb{R}$. It may be surprising that although we use the Bohr measure for $\varphi(x_I)$, there appears momentum-like variable with Lebesgue measure. We treat it here as an auxiliary variable that is used in the perturbative expression for an exponent of quantum scalar constraint operator in order to put it in a spin-foam form. This indicates that there may be another approach where the Lebesgue measure for $\varphi(x_I)$ is used, we limit however to the standard polymer quantization. 
\subsection{Foams}\label{sc:foams}
The operator $\CE[x_I]$ changes the graph by adding or subtracting a loop tangential to two different links of the graph $\gamma$ at the node $x_I$. An admissible history of an initial graph $\gamma_i$ into a final graph $\gamma_f$ is a 2-complex $\kappa$ embedded in $\spacetime=\slice\times \interval$ such that 
\begin{itemize}
\item its intersection with each slice $\slice_t=\slice\times\{t\}$, $t\in ]0,1]$, is a graph $\gamma_t$ of the following form:
$$
\gamma_t = \Gamma \cup \loops_t,
$$
where $\loops_t$ is a set of all loops (marked circles embedded in $\slice_t$ which marked points coincide with a nodes of $\Gamma$) tangential to two non-tangential links of $\Gamma$ (we assume that different loops tangential to the same pair of links differ by the maximal order of tangentiality to the links).
\item $\gamma_0=\gamma^\dagger_\inn$, where $\gamma^\dagger_\inn$ is the graph obtained from $\gamma_\inn$ by flipping orientation of each of its links,
\item $\gamma_1 \sim f(\gamma_\out)$ for some diffeomorphism $f\in \Diff$ (let us recall that $\gamma_1 \sim f(\gamma_\out)$ means that the two graphs are equal up to the operations of flipping orientations of links, splitting links, adding links and nodes -- see the beginning of Section \ref{sc:scalar_constraint}).
\end{itemize}

Let us note that the $2$-complex $\kappa$ contains a subcomplex $\Gamma\times \interval$. An example of an admissible 2-complex is depicted on figure \ref{fig:Example_1}. 
We choose an orientation of each internal edge $e=\{(x_I,t): t\in [a,b]\}$ such that $(x_I,a)$ is its beginning and $(x_I,b)$ is its end. Each face is oriented in such a way that for each $t\in ]0,1]$ the orientation of the faces of the foam $\complex\cap ( \slice\times [0,t])$ agree with the orientation of $\gamma_t$.

\subsection{Spin-foam operator}
We aim at giving a graphical calculus for calculating \eqref{eq:sfoperator} that amounts to sum over spin-foam amplitudes. First, we will describe an operator spin-foam formulation \cite{OSM,KisielowskiPhD}. In this formulation the faces of the foam are labelled with irreducible representations of the SU(2) group, edges with certain operators and vertices with certain contractors.

The faces of $\kappa$ of the form $f=\link\times \interval$, where $\link$ is a link of $\Gamma$, are labelled with representation $\rho_f=\rho_\link$. All other faces $f'$ are labelled with representation $\rho_{f'}=\rho_{(l)}$, where $\rho_{(l)}$ is an irreducible representation of SU(2) with spin $l$.

Denote by $\R_e$ the sequence of representations corresponding to the edge $e$, i.e.
$$
\R_e=(\rho_{f_1},\ldots,\rho_{f_M},\rho_{f_{M+1}}^*,\ldots, \rho_{f_N}^*),
$$
where $f_1,\ldots, f_M$ are the faces intersecting the edge $e$ and their orientations agree with the orientation of $e$, $f_{M+1},\ldots,f_{N}$ are faces intersecting $e$ which orientation is opposite to the orientation of $e$. Let $\varepsilon_e$ be defined by
$$\varepsilon_{rs}=\cases{0,&if $f_r$ and $f_s$ are tangential at the edge $e$,\\
1&otherwise.\\}$$
Let $e=x_I\times[a,b]$ be an edge. We will denote by $s_e$ its source and by $t_e$ its target. We also introduce $p_e$ by definition equal to $p_{x_I}$ and $\sign_e$ by definition equal to $\sign_{x_I}$. We label each edge with an operator $P_e^\sign$:
$$
P_e^\sign:= e^{-\iu p_{e} (\varphi(t_e)-\varphi(s_e))} (p^2_{e}-\CL[(\R_e,\varepsilon_e)]+\iu\epsilon\sign_e)^{-1},
$$
where $\varphi:\spacetime\to \mathbb{R}$ such that 
$$\varphi(x)=\varphi_\inn(x) {\rm\ for\ }x\in \nodes{\gamma_0},\quad \varphi(x)=\varphi_\out(x) {\rm\ for\ }x\in \nodes{\gamma_1}.$$

Each internal vertex $v$ is labeled with a contractor $\contractor_v$. As in \cite{KisielowskiPhD} we introduce a vertex Hilbert space
$$
\Hil_v=\bigotimes_{e{\rm\ incoming\ at\ }v}\Hil_{\R_e}\otimes \bigotimes_{e'{\rm\ outgoing\ from\ }v}\Hil_{\R_{e'}}^*.
$$
A contractor $\contractor_v$ is a linear functional:
$$
\contractor_v\in \Hil_v^*.
$$
In our case there are just 2 edges incident at each vertex $v$, one incoming $e_v$ and one outgoing $e'_v$. There are two cases:
\begin{enumerate}
\item There is a face $f_{rs}$ tangent to the faces $f_r$ and $f_s$ that contains $e'_v$ but does not contain $e_v$. In this case:
$$
\contractor_v=\CE[(\R_e,rs)]^\dagger.
$$
\item There is a face $f_{rs}$ tangent to the faces $f_r$ and $f_s$ that contains $e_v$ but does not contain $e'_v$. In this case:
$$
\contractor_v=\frac{1}{2l+1}\CE[(\R_e,rs)].
$$
The factor $\frac{1}{2l+1}$ comes from the fact that the scalar product in the spin-network basis is given by the Haar integrals over the SU(2) group (see \eqref{eq:CEadj}).
\end{enumerate}
The canonical contraction is \cite{OSM, KisielowskiPhD}:
$$
\Tr\left(\complex,\rho,P,\contractor\right)=\bigotimes_{v\in \complex^{(0)}} \contractor_v \lrcorner \bigotimes_{e\in \complex^{(1)}} P_e^\sign.
$$
Using this contraction the amplitude $A([s_f],\varphi_f(x_I);[s_i],\varphi_i(x_I)) $ can be written in the following form:
$$
A([s_f],\varphi_f(x_I);[s_i],\varphi_i(x_I)) =\lim_{\epsilon\to 0^+}\sum_{M=0}^{\infty} \lambda^M\,  A_M([s_f],\varphi_f(x_I);[s_i],\varphi_i(x_I)),
$$
where
\begin{eqnarray} 
\fl A_M([s_f],\varphi_f(x_I);[s_i],\varphi_i(x_I))=\\ \fl =  \frac{1}{\NV}\sum_{\sign} \sum_{\sigma\in S_n}\sum_{\kappa_M} \frac{1}{(\pi\iu)^n}\prod_{I=1}^n\int_{-\infty}^{+\infty}  dp_{x_{I}} p_{x_{I}} \theta(p_{x_{I}}) \sign_{x_I} \brakett{[s_\out] \left| U_\sigma^{\rm gr} \Tr\left(\complex,\rho,P^\sign,\contractor\right)\right|[s_\inn]},\label{eq:sfoperatorI}
\end{eqnarray}
$\kappa_M$ ranges all spin foams in our class that have $M$ internal vertices and a boundary defined by graphs $\gamma_\inn$ and $\gamma_\out$.
\subsection{Spin-foam amplitudes}

In order to pass from the spin-foam operator to spin-foam amplitudes, we insert decompositions of identity in terms of the eigenvalue bases of the operators $\CL[(\R_e,\varepsilon_e)]$. The result of this procedure can be summarized by the following prescription for the spin foams and spin-foam amplitudes.

A spin foam $\spinfoam=(\kappa,\rho,\iota,p,\sign)$ is a complex $\kappa$ in the class described in section \ref{sc:foams} together with a coloring of its faces with unitary irreducible representations of the SU(2) group (as described in the previous section) and edges with triples $(\iota_{e},p_{e},\sign_{e})$, where $\iota_e$ is an eigenvector of the operator $\CL[(\R_e,\varepsilon_e)]$ and $p_{e}$ is a real number (the scalar field "momentum") and $\sign_{e}$ is a number equal to $+1$ or $-1$. We assume that $p_{e}=p_{e'}=p_x,\,\sign_{e}=\sign_{e'}=\sign_x$ if $e,e'\subset x\times [0,1]$ for some node $x\in\nodes{\gamma_0}$. 

We will denote by $\eCL[e]$ the eigenvalue of the operator $\CL[(\R_e,\varepsilon_e)]$. To each edge $e$ we assign an edge amplitude:
$$
\Amplitude{e}=\frac{e^{-\iu p_{e} (\varphi(t_e)-\varphi(s_e))}}{p_e^2-\eCL[e]+ \iu \sign(e) \epsilon},
$$
where $t_e$ is the target of $e$ and $s_e$ is its source.

To each internal vertex $v$ we assign a vertex amplitude, which is equal to
$$\Amplitude{v}=\contractor_v(\iota_{e_v}\otimes \iota_{e'_v}^\dagger),$$
where $e_v$ is the edge incoming to $v$ and $e'_v$ is the edge outgoing from $v$. There are two cases:
\begin{enumerate}
\item There is a face $f_{rs}$ tangent to the faces $f_r$ and $f_s$ that contains $e'_v$ but does not contain $e_v$, i.e a loop is created at the vertex $v$. In this case:
$$
\Amplitude{v}=\braket{\iota_{e'_v}}{\CE[(\R_{e_v},rs)]^\dagger \iota_{e_v}}.
$$
\item There is a face $f_{rs}$ tangent to the faces $f_r$ and $f_s$ that contains $e_v$ but does not contain $e'_v$, i.e a loop is annihilated at the vertex $v$. In this case:
$$
\contractor_v=\frac{1}{2l+1} \braket{\iota_{e'_v}}{\CE[(\R_{e_{v}},rs)] \iota_{e_v}}.
$$
\end{enumerate}

Finally, to each spin foam $\spinfoam=(\complex,\rho,\iota,p,\sign)$ we assign a spin-foam amplitude
$$
\Amplitude{\spinfoam}=\prod_{e\in \complex^{(1)}} \Amplitude{e} \prod_{v\in \complex^{(0)}} \Amplitude{v}
$$ 
The boundary of the foam is formed from 2 disjoint graphs. Therefore the spin-network state induced on the boundary (defined in \cite{SFLQG}) is a tensor product of of two spin-network states:
$$
\ket[s_{\rm boundary}(\spinfoam)]=\ket[s_1({\spinfoam})]\otimes\ket[s_0(\spinfoam)^\dagger],
$$
In the formula above $s^\dagger$ denotes the spin network conjugate to $s$ (for the definition we refer our reader to \cite{SFLQG}). The amplitude $A([s_\out],\varphi_\out(x_I);[s_\inn],\varphi_\inn(x_I)) $ can be expressed in the power series of $\lambda$:
$$A([s_\out],\varphi_\out(x_I);[s_\inn],\varphi_\inn(x_I)) =: \sum_{M=0}^{\infty} \lambda^M A_M([s_\out],\varphi_\out(x_I);[s_\inn],\varphi_\inn(x_I)).$$
Each coefficient $A_M([s_\out],\varphi_\out(x_I);[s_\inn],\varphi_\inn(x_I))$ in this expansion can be written as a sum over spin foams and residual diffeomorphisms (we assume that the sum over $M$ and the limit can be interchanged):
$$
A_M([s_\out],\varphi_\out(x_I);[s_\inn],\varphi_\inn(x_I))=\lim_{\epsilon\to 0^+}\frac{1}{\NV} \sum_{\sigma\in S_n} \sum_{F_M} \brakett{[s_\out] \right| U^{\rm gr}_{\sigma} \left| [s_1(\spinfoam_M)]} \Amplitude{\spinfoam_M}  \braket{[s_0(\spinfoam_M)]}{[s_\inn]},
$$
where 
\begin{eqnarray*}
\sum_{F_M} \brakett{[s_\out] \right| U^{\rm gr}_{\sigma} \left| [s_1(\spinfoam_M)]} \Amplitude{\spinfoam_M}  \braket{[s_0(\spinfoam_M)]}{[s_\inn]}=\\ \fl \sum_{\kappa_M} \sum_{\iota_e} \sum_{\sign} \frac{1}{(\pi\iu)^n}\prod_{I=1}^n\int_{-\infty}^{+\infty}  dp_{x_{I}} p_{x_{I}} \theta(p_{x_{I}}) \sign_{x_I}\  \Amplitude{\spinfoam_M} \brakett{[s_\out] \right| U^{\rm gr}_{\sigma} \left| [s_1(\spinfoam_M)]} \braket{[s_0(\spinfoam_M)]}{[s_\inn]},
\end{eqnarray*}
$\kappa_M$ ranges over all foams with $M$ internal vertices and boundary formed by graphs defined by $\gamma_\inn$ and $\gamma_\out$, $\iota_e$ ranges over an orthonormal basis of eigenvectors of $\CL[(\R_e,\varepsilon_e)]$. Let us notice that in contrast to the standard spin-foam sums, there is no sum over the spins because they are fixed by the spins on the boundary spin networks at all orders.

In the case of massless scalar field the integral over the momenta and the limit of $\epsilon$ going to $0^+$ can be performed explicitly using the technique described in \ref{sc:Appendix_A}. Therefore it is convenient to introduce also the following amplitude:
$$
\Amplitude{(\complex,\rho,\iota)}:= \lim_{\epsilon\to 0^+} \sum_{\sign} \frac{1}{(\pi\iu)^n}\prod_{I=1}^n\int_{-\infty}^{+\infty}  dp_{x_{I}} p_{x_{I}} \theta(p_{x_{I}})  \Amplitude{(\complex,\rho,\iota,p,\sign)}.
$$
Since $s_1(\spinfoam)$ and $s_0(\spinfoam)$ depend neither on the matter degrees of freedom $p$ nor on $\sign$, the sum over spin foams can be reduced to a sum over gravitational degrees of freedom $\spinfoam^{\rm gr}=(\complex,\rho,\iota)$ leading to the following expression:
\begin{eqnarray*}
A_M([s_\out],\varphi_\out(x_I);[s_\inn],\varphi_\inn(x_I))=\\=\frac{1}{\NV} \sum_{\sigma\in S_n} \sum_{\complex_M} \sum_{\iota_e} \brakett{[s_\out] \right| U^{\rm gr}_{\sigma} \left| [s_1(\complex_M,\rho,\iota)]} \Amplitude{(\complex_M,\rho,\iota)}  \braket{[s_0(\complex_M,\rho,\iota)]}{[s_\inn]}.
\end{eqnarray*}
As will be shown explicitly by the example in the next section, the amplitude $\Amplitude{(\complex_M,\rho,\iota)}$ (in contrast to $\Amplitude{(\complex_M,\rho,\iota,p,\sign)}$) does not have a local form, i.e. it cannot be written as product of $\prod_e \Amplitude{e} \prod_v \Amplitude{v}$, where $\Amplitude{e}$ depends only on the edge $e$ and $\Amplitude{v}$ depends only on the vertex $v$. Similar issue has been encountered in the symmetry reduced models and solved by introducing a regulator ($\epsilon$) \cite{Localsfexpansion} (see also Section \ref{sc:ACHHRVW}). Therefore it is not surprising that after removing the regulator the amplitude no longer has a local form.

\section{Example}\label{sc:Example}
\begin{figure}
\begin{subfigure}{0.45\textwidth}
 \centering
\includegraphics[width=.9\linewidth]{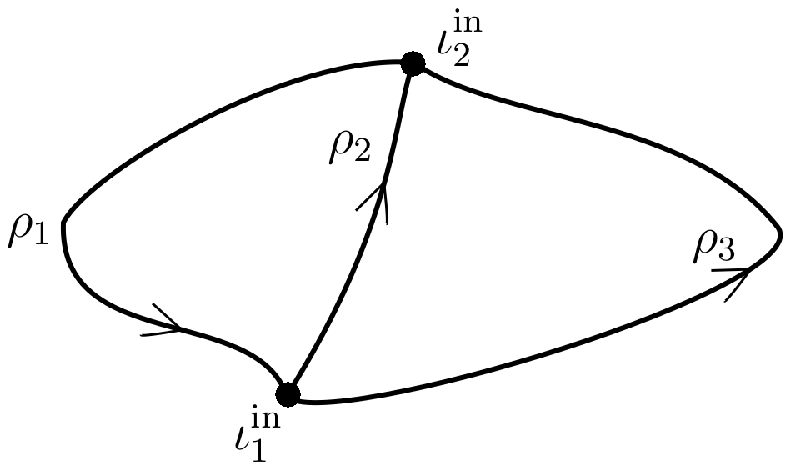}
  \caption{The spin network $s_\inn$.}
  \label{fig:Example_1_sni}
\end{subfigure}
\begin{subfigure}{0.45\textwidth}
 \centering
\includegraphics[width=.9\linewidth]{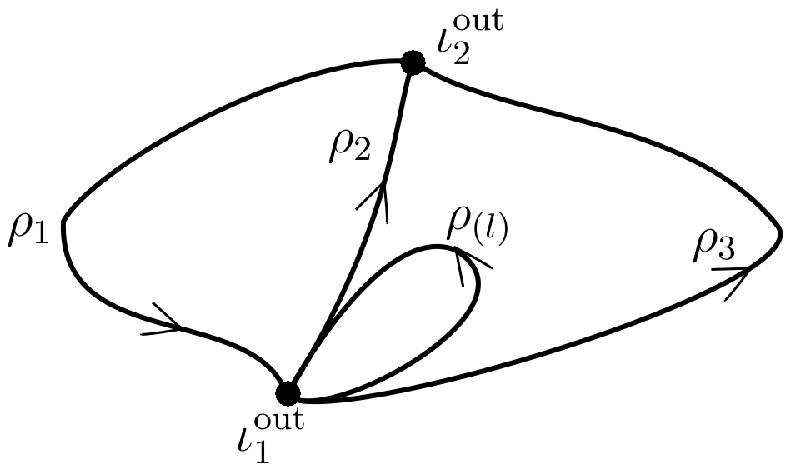}
  \caption{The spin network $s_\out$.}
  \label{fig:Example_1_snf}
\end{subfigure}
\caption{Example 1: We consider a physical scalar product between states such that the quantum gravitational degrees of freedom are encoded in the spin-network states $\ket[{[s_\inn]}]$ and $\ket[{[s_\out]}]$. We will assume that $\rho_1\not\sim \rho_2$, $\rho_2\not\sim \rho_3$, $\rho_1\not\sim \rho_3$.}
\label{fig:Example_1_sn}
\end{figure}

As an example we consider a physical scalar product between states 
$$\ket[\Psi_{\inn\slash\out}] = \ket[{[s_{\inn\slash \out}]}] \otimes \ket[\Psi^{\rm mat}_{\inn\slash \out}],$$
where the spin networks graphs $s_1$ and $s_2$ are depicted on figure \ref{fig:Example_1_sn}. We will assume that $\rho_1\not\sim \rho_2$, $\rho_2\not\sim \rho_3$, $\rho_1\not\sim \rho_3$ (see \eqref{eq:simplification}).

Let us also assume that at each node $x_I$ of each of the graphs $\gamma_\inn, \gamma_\out$ the tangent vectors to links meeting at $x_I$ span the whole tangent space at $x_I$. With this assumption the diffeomorphisms $f\in \Diff_{\nodes{\gamma}}\slash \TDiff{\gamma}$ such that $f(\gamma)=\gamma$ ($\gamma=\gamma_\inn$ or $\gamma=\gamma_\out$) coincide with the permutations of the links of $\gamma$.
\subsection{The zeroth order}
In the zeroth order there are no foams that contribute:
$$
A_0([s_\out],\varphi_\out(x_I);[s_\inn],\varphi_\inn(x_I))=0.
$$
\subsection{The first order}
\begin{figure}
\begin{subfigure}{0.45\textwidth}
 \centering
\includegraphics[width=.9\linewidth]{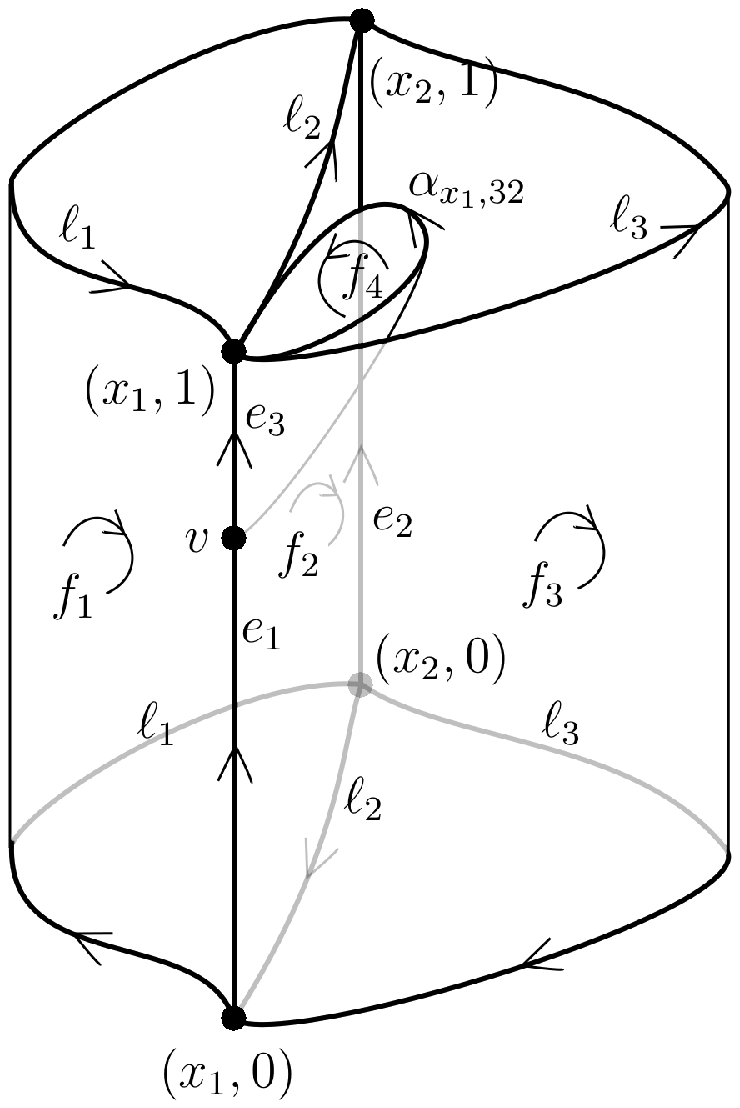}
  \caption{The foam $\kappa_1^1$. There is another foam $\kappa_1^2$ that adds a loop $\alpha_{x_1, 23}$, which differs from the loop $\alpha_{x_1, 32}$ only by orientation.}
  \label{fig:Example_1a}
\end{subfigure}
\begin{subfigure}{0.45\textwidth}
 \centering
\includegraphics[width=.9\linewidth]{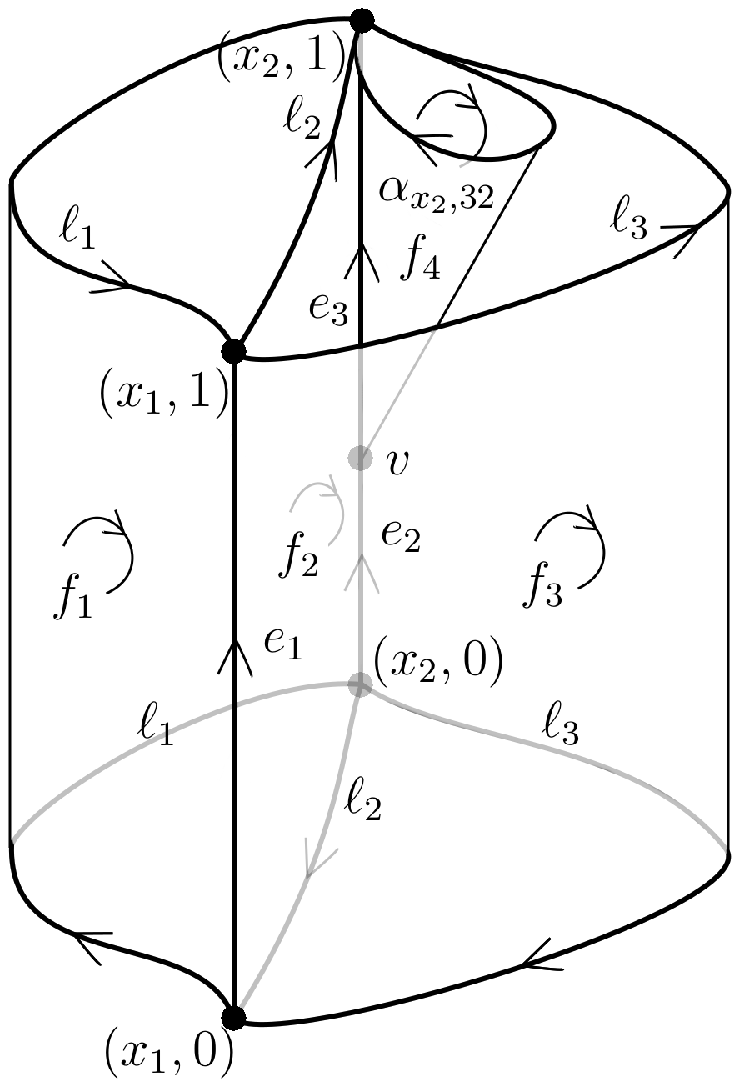}
  \caption{The foam $\kappa_1^3$. There is another foam $\kappa_1^4$ that adds a loop $\alpha_{x_2, 23}$, which differs from the loop $\alpha_{x_2, 32}$ only by orientation.}
  \label{fig:Example_1b}
\end{subfigure}
\caption{Example 1: In the first order of the (vertex) expansion four foams contribute. The first pair $\kappa_1^1,\kappa_1^2$ represents histories of a graph in which a loop is added at the node $x_1$ between links $\link_2$ and $\link_3$ and the second pair $\kappa_1^3,\kappa_1^4$ represents histories in which a loop is added at $x_2$ between the same links. In the first case the diffeomorphism bringing the final graph of the foam into the graph $\gamma_2$ is a trivial one and in the second case the diffeomorphism transposes the two nodes $x_1$ and $x_2$. Let us notice that foams adding loops between $\link_1$ and $\link_2$ or $\link_1$ and $\link_3$ give zero contribution due to our assumption that $\rho_1\not \sim \rho_2$, $\rho_1\not \sim \rho_3$, $\rho_2\not \sim \rho_3$.}
\label{fig:Example_1}
\end{figure}

In the first order there are four foams that contribute (let us recall that each graph represents an equivalence class of graphs modulo diffeomorphisms fixing each node of the graph). They are depicted on figure \ref{fig:Example_1}.

\subsubsection{Foam $\kappa_1^1$}
Let us first focus on the foam $\kappa_1^1$ from figure \ref{fig:Example_1a}. There is only one (up to equivalence of representations) possible coloring of the faces. The coloring of the faces $f_1,\,f_2,\,f_3$ is fixed by the coloring of the boundary links:
$$
\rho_{f_1}=\rho_1,\quad \rho_{f_2}=\rho_2,\quad \rho_{f_3}=\rho_3.
$$
The face $f_4$ is labelled with the chosen representation $\rho_{(l)}$:
$$
\rho_{f_4}=\rho_{(l)}.
$$
With the edge $e_1$ there is associated a sequence of representations $\R_{e_1}=(\rho_1^*,\rho_2,\rho_3)$ and a symmetric function $\epsilon$ that is equal to 1 for any pair of indices $r,s$ labelling the faces intersecting $e_1$. It is labelled with a pair $(p_{e_1},\iota_{e_1})$:
\begin{itemize}
\item a real number $p_{e_1}=p_{x_1}$,
\item an invariant tensor $\iota_{e_1}\in \Inv{\Hil_{\rho_1}^*\otimes \Hil_{\rho_2}\otimes \Hil_{\rho_3}}$ that is an eigenvector of the operator $\CL[(\R_{e_1},\epsilon_{e_1})]$ with eigenvalue $\CL[e_1]$.
\end{itemize}
With the edge $e_2$ there is associated a sequence of representations $\R_{e_2}=(\rho_1,\rho_2^*,\rho_3^*)$ and a symmetric function $\epsilon$ that is equal to 1 for any pair of indices $r,s$ labelling the faces intersecting $e_1$. It is labelled with a pair $(p_{e_2},\iota_{e_2})$:
\begin{itemize}
\item a real number $p_{e_2}=p_{x_2}$,
\item an invariant tensor $\iota_{e_2}\in \Inv{\Hil_{\rho_1}\otimes \Hil_{\rho_2}^*\otimes \Hil_{\rho_3}^*}$ that is an eigenvector of the operator $\CL[(\R_{e_2},\epsilon_{e_2})]$ with eigenvalue $\CL[e_2]$.
\end{itemize}
With the edge $e_3$ there is associated a sequence of faces $(f_1,f_2,f_4,f_4,f_3)$ and representations $\R_{e_3}=(\rho_1^*,\rho_2,\rho_{(l)}^*, \rho_{(l)}, \rho_3)$. The function $\epsilon$ is:
$$
\epsilon_{e_3,\,rs}=\cases{0,&if $r=2, s=3$ or  $r=3, s=2$,\\
0,&if $r=4, s=5$ or  $r=5, s=4$,\\
1&otherwise.\\}
$$
It is labelled with a pair $(p_{e_3},\iota_{e_3})$:
\begin{itemize}
\item a real number $p_{e_3}=p_{x_1}$,
\item an invariant tensor $\iota_{e_3}\in \Inv{\Hil_{\rho_1}^*\otimes \Hil_{\rho_2}\otimes \Hil_{\rho_{(l)}}^*\otimes \Hil_{\rho_{(l)}} \otimes \Hil_{\rho_3}}$ that is an eigenvector of the operator $\CL[(\R_{e_3},\epsilon_{e_3})]$ with eigenvalue $\CL[e_3]$.
\end{itemize}

The edge amplitudes are the following:
$$
\Amplitude{e_1}=\frac{e^{-\iu p_{e_1} (\varphi(v)-\varphi((x_1,0)))}}{p_{e_1}^2-\eCL[e_1]+ \iu \sign(x_1) \epsilon},\quad \Amplitude{e_2}^{\sign}=\frac{e^{-\iu p_{e_2} (\varphi((x_2,1))-\varphi((x_2,0)))}}{p_{e_2}^2-\eCL[e_2]+ \iu \sign(x_2) \epsilon},\quad
\Amplitude{e_3}=\frac{e^{-\iu p_{e_3} (\varphi((x_1,1))-\varphi(v))}}{p_{e_3}^2-\eCL[e_3]+ \iu \sign(x_1) \epsilon}.
$$
The vertex amplitude is:
$$
\Amplitude{v}=\braket{\iota_{e_3}}{\CE[(\R_{e_1},23)]^\dagger \iota_{e_1}}.
$$
The spin-foam amplitude is
\begin{eqnarray*}
\Amplitude{\spinfoam_1^1} = \prod_e \Amplitude{e} \prod_v \Amplitude{v} =\\ = \frac{e^{-\iu p_{x_1} (\varphi_\out(x_1)-\varphi_\inn(x_1))-\iu p_{x_2} (\varphi_{\out}(x_2)-\varphi_{\inn}(x_2))}}{(p_{x_1}^2-\eCL[e_1]+ \iu \sign(x_1) \epsilon)(p_{x_1}^2-\eCL[e_3]+ \iu \sign(x_1) \epsilon)(p_{x_2}^2-\eCL[e_2]+ \iu \sign(x_2) \epsilon)}  \braket{\iota_{e_3}}{\CE[(\R_{e_1},32)]^\dagger \iota_{e_1}}.
\end{eqnarray*}
Using the contour technique studied in \ref{sc:Appendix_A} we can integrate over the scalar field momenta:
$$
\Amplitude{(\kappa_1^1,\rho,\iota)} =\lim_{\epsilon\to 0^+} \frac{1}{(\pi \iu)^2}\sum_{\sign} \int_{-\infty}^{+\infty}  dp_{x_{1}} p_{x_{1}} \theta(p_{x_{1}}) \int_{-\infty}^{+\infty}  dp_{x_{2}} p_{x_{2}} \theta(p_{x_{2}}) \sign_{x_1} \sign_{x_2} \Amplitude{(\kappa_1^1,\rho,\iota,p,\sign)}.
$$
There are 2 cases:
\begin{enumerate}
\item $\eCL[e_1]=\eCL[e_3]$: 
$$
\Amplitude{(\kappa_1^1,\rho,\iota)} = \braket{\iota_{e_3}}{\CE[(\R_{e_1},32)]^\dagger \iota_{e_1}} \frac{d}{d \eCL[e_1]} e^{-\iu \sqrt{\eCL[e_1]} (\varphi((x_1,1))-\varphi((x_1,0)))-\iu \sqrt{\eCL[e_2]} (\varphi((x_2,1))-\varphi((x_2,0)))},
$$
\item $\eCL[e_1]\neq\eCL[e_3]$:
\begin{eqnarray*}
\Amplitude{(\kappa_1^1,\rho,\iota)} = \braket{\iota_{e_3}}{\CE[(\R_{e_1},32)]^\dagger \iota_{e_1}} \frac{e^{-\iu \sqrt{\eCL[e_1]} (\varphi((x_1,1))-\varphi((x_1,0)))-\iu \sqrt{\eCL[e_2]} (\varphi((x_2,1))-\varphi((x_2,0)))}}{\eCL[e_1]-\eCL[e_3]}+\\+ \braket{\iota_{e_3}}{\CE[(\R_{e_1},32)]^\dagger \iota_{e_1}}\frac{e^{-\iu \sqrt{\eCL[e_3]} (\varphi((x_1,1))-\varphi((x_1,0)))-\iu \sqrt{\eCL[e_2]} (\varphi((x_2,1))-\varphi((x_2,0)))}}{\eCL[e_3]-\eCL[e_1]}.
\end{eqnarray*}
\end{enumerate}
Since we assumed that $\rho_1\not \sim \rho_2$, $\rho_1\not \sim \rho_3$, $\rho_2\not\sim \rho_3$ there is only one nonzero term corresponding to the identity diffeomorphism $f\in \Diff_{\nodes{\gamma}}\slash \TDiff{\gamma}$ that according to our formula \eqref{eq:partialdifffix} contributes to the scalar product $\braket{[s_0(\spinfoam^1_1)]}{[s_\inn]}$:
$$
\braket{[s_0(\spinfoam^1_1)]}{[s_\inn]}= \braket{s_0(\spinfoam^1_1)}{s_\inn},
$$
where $N_{\gamma_0}=6$ is the number of permutations of the links of the theta graph. The diffeomorphism bringing the upper boundary of the foam into the graph $\gamma_\out$ is the identity diffeomorphism. Therefore 
$$
\brakett{[s_\out] \right| U^{\rm gr}_{\sigma} \left| [s_1(\spinfoam_1^1)]} = \braket{[s_\out] }{ [s_1(\spinfoam_1^1)]} =  \braket{s_\out }{s_1(\spinfoam_1^1)}.
$$
\subsubsection{Foam $\kappa_1^2$}
For the foam $\kappa_1^2$ the formulas are the same as for $\kappa_1^1$ except that $\CE[(\R_{e_1},32)]^\dagger$ is replaced by $\CE[(\R_{e_1},23)]^\dagger$ and 
$$
\brakett{[s_\out] \right| U^{\rm gr}_{\sigma} \left| [s_1(\spinfoam_1^2)]} = \braket{[s_\out] }{ [s_1(\spinfoam_1^2)]} =\ \braket{s_\out }{s_1(\spinfoam_1^2)}= \braket{s_\out }{\tilde{s}_1(\spinfoam_1^2)},
$$
where $\tilde{s}_1(\spinfoam_1^2)$ is the spin network obtained from $s_1(\spinfoam_1^2)$ by flipping the orientation of the loop $\alpha_{x_1, 23}$.
\subsubsection{Foam $\kappa_1^3$}
\begin{figure}
\begin{subfigure}{0.45\textwidth}
 \centering
\includegraphics[width=.9\linewidth]{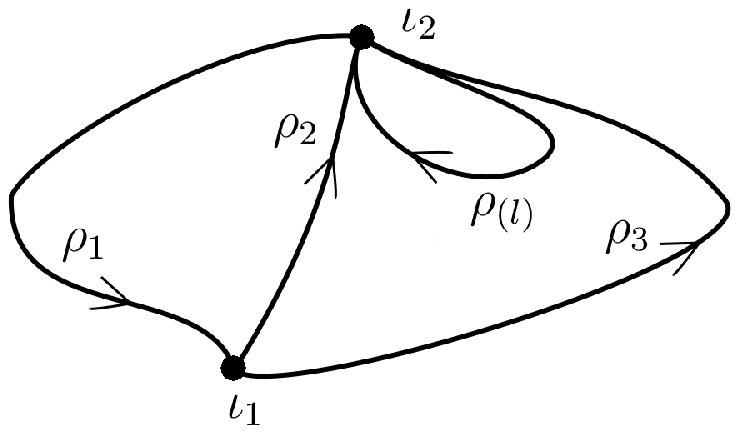}
  \caption{The spin network $s_1(\spinfoam_1^3)$.}
  \label{fig:s1}
\end{subfigure}
\begin{subfigure}{0.45\textwidth}
 \centering
\includegraphics[width=.9\linewidth]{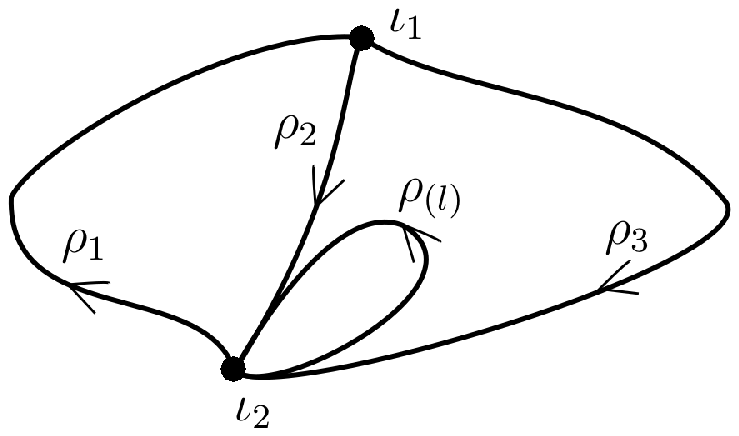}
  \caption{The spin network $s^{(12)}_1(\spinfoam_1^3)$.}
  \label{fig:s1f}
\end{subfigure}
\centering
\begin{subfigure}{0.45\textwidth}
\includegraphics[width=.9\linewidth]{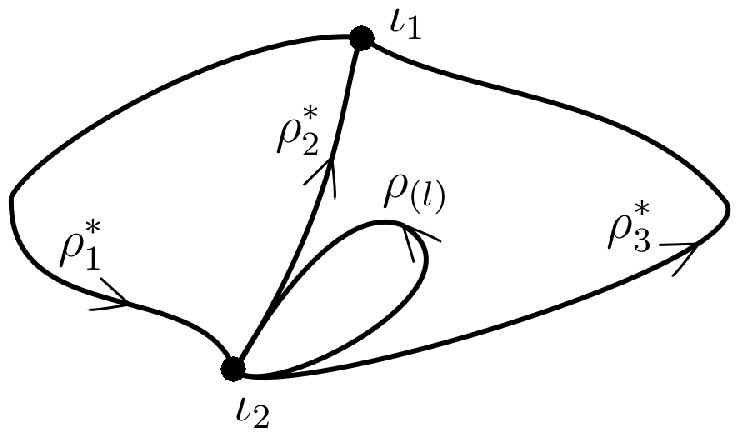}
  \caption{The spin network $\tilde{s}^{{(12)}}_1(\spinfoam_1^3
  )$}
  \label{fig:tildesf}
\end{subfigure}
\caption{The diffeomorphism $\sigma$ bringing the graph $\gamma_1(\complex_1^3)$ into the graph $\gamma_\out$ transposes the nodes $x_1$ and $x_2$: $\sigma=(12)$. The result of the action of this diffeomorphism on $s_1(\spinfoam_1^3)$ is denoted by $s^{{(12)}}_1(\spinfoam_1^3)$. In order to evaluate the scalar product $\braket{[s_\out] }{ [s^{(12)}_1(\spinfoam_1^3)]}$ we consider an equivalent spin network $\tilde{s}^{{(12)}}_1(\spinfoam_1^3)$ obtained from $s^{(12)}_1(\spinfoam_1^3)$ by operations of flipping the orientations of the links $\link_1,\, \link_2, \, \link_3$. As a result $\brakett{[s_\out] \right| U^{\rm gr}_{\sigma} \left| [s_1(\spinfoam_1^3)]} = \braket{s_\out}{\tilde{s}^{{(12)}}_1(\spinfoam_1^3)}$. The scalar product $\braket{s_\out}{\tilde{s}^{{(12)}}_1(\spinfoam_1^3)}$ can be evaluated using formula \eqref{eq:snscalarproduct}.}
\label{fig:final_graph}
\end{figure}

The calculation for the second foam $\kappa_1^3$ is also completely analogous and leads to the following 2 cases:
\begin{enumerate}
\item $\eCL[e_2]=\eCL[e_3]$: 
$$
\Amplitude{(\kappa_1^3,\rho,\iota)} = \braket{\iota_{e_3}}{\CE[(\R_{e_2},23)]^\dagger \iota_{e_2}} \frac{d}{d \eCL[e_2]} e^{-\iu \sqrt{\eCL[e_2]} (\varphi_\out(x_2)-\varphi_\inn(x_2))-\iu \sqrt{\eCL[e_1]} (\varphi_\out(x_1)-\varphi_\inn(x_1))},
$$
\item $\eCL[e_2]\neq\eCL[e_3]$:
\begin{eqnarray*}
\Amplitude{(\kappa_1^3,\rho,\iota)} = \braket{\iota_{e_3}}{\CE[(\R_{e_2},23)]^\dagger \iota_{e_2}} \frac{e^{-\iu \sqrt{\eCL[e_2]} (\varphi_\out(x_2)-\varphi_\inn(x_2))-\iu \sqrt{\eCL[e_1]} (\varphi_\out(x_1)-\varphi_\inn(x_1))}}{\eCL[e_2]-\eCL[e_3]}+\\+ \braket{\iota_{e_3}}{\CE[(\R_{e_2},23)]^\dagger \iota_{e_2}}\frac{e^{-\iu \sqrt{\eCL[e_3]} (\varphi_\out(x_2)-\varphi_\inn(x_2))-\iu \sqrt{\eCL[e_1]} (\varphi_\out(x_1)-\varphi_\inn(x_1))}}{\eCL[e_3]-\eCL[e_2]}.
\end{eqnarray*}
\end{enumerate}
The diffeomorphism $\sigma$ bringing the graph $\gamma_1(\complex_1^3)$ into $\gamma_\out$ transposes the two nodes $x_1$ and $x_2$ (see figure \ref{fig:final_graph}), i.e as a permutation:
$$
\sigma=(12).
$$
Let us denote by $s_1^{(12)}(\spinfoam_1^3)$ the spin network obtained from $s_1(\spinfoam_1^3)$ by the action of $\sigma=(12)$. In this case
$$
\brakett{[s_\out] \right| U^{\rm gr}_{\sigma} \left| [s_1(\spinfoam_1^3)]} = \braket{[s_\out] }{ [s^{(12)}_1(\spinfoam_1^3)]}=\braket{s_\out}{\tilde{s}^{{(12)}}_1(\spinfoam_1^3)},
$$
where $\tilde{s}^{{(12)}}_1(\spinfoam_1^3)$ is the spin network obtained from $s^{{(12)}}_1(\spinfoam_1^3)$ by operations of flipping the orientations of the links $\link_1,\, \link_2, \, \link_3$.

\subsubsection{Foam $\kappa_1^4$}
For the foam $\kappa_1^4$ the formulas are the same as for $\kappa_1^3$ except that $\CE[(\R_{e_2},32)]^\dagger$ is replaced by $\CE[(\R_{e_2},23)]^\dagger$ and 
$$
\brakett{[s_\out] \right| U^{\rm gr}_{\sigma} \left| [s_1(\spinfoam_1^4)]} = \braket{s_\out }{\dbtilde{s}^{(12)}_1(\spinfoam_1^4)},
$$
where $\dbtilde{s}^{(12)}_1(\spinfoam_1^4)$ is the spin network obtained from $s^{(12)}_1(\spinfoam_1^4)$ by flipping the orientation of $\link_1,\, \link_2, \, \link_3$ and the loop $\alpha_{x_1, 23}$.
\subsection{The physical scalar product}
In the second order, there are no foams that contribute. The next non-trivial order is the third order. Therefore the physical scalar product can be written in the following form:
\begin{eqnarray*}
\fl \phys{\Psi_\out}{\Psi_\inn}=\int \prod_{I=1}^{n}\Bohr[\varphi_\out(x_I)] \Bohr[\varphi_\inn(x_I)] \overline{\Psi_\out^{\rm mat}}(\varphi_\out)\Psi_\inn^{\rm mat}(\varphi_\inn)\cdot \\ \fl \cdot \frac{\lambda}{2!} (\sum_{\iota_e} \braket{s_\out }{s_1(\spinfoam_1^1)} \Amplitude{(\complex_1^1,\rho,\iota)}\braket{s_0(\spinfoam^1_1)}{s_\inn}+\sum_{\iota_e}\braket{s_\out }{\tilde{s}_1(\spinfoam_1^2)}\Amplitude{(\complex_1^2,\rho,\iota)}\braket{s_0(\spinfoam^1_2)}{s_\inn}+\\ \fl +\sum_{\iota_e}\braket{s_\out}{\tilde{s}^{{(12)}}_1(\spinfoam_1^3)} \Amplitude{(\complex_1^3,\rho,\iota)} \braket{s_0(\spinfoam^3_1)}{s_\inn}+\sum_{\iota_e}\braket{s_\out }{\dbtilde{s}^{(12)}_1(\spinfoam_1^4)}\Amplitude{(\complex_1^4,\rho,\iota)} \braket{s_0(\spinfoam^4_1)}{s_\inn})+\mathcal{O}(\lambda^3).
\end{eqnarray*}
Each of the spin-network scalar products can be directly evaluated using formula \eqref{eq:snscalarproduct}.

\section{Discussion and Outlook}
The existence of the link between the canonical Loop Quantum Gravity and covariant spin-foam theory has been long debated \cite{Baezstringsloops,Iwasaki,R, RR, RovelliProjector, Markopoulou, BaezSF, NouiPerez3d, DanieleQGroup, EPRL, ETHCorrespondenceI, ETHCorrespondenceII,ETHCorrespondenceIII,DittrichCorrespondence,ATZCorrespondence,ThiemannZipfel, ZipfelPhD, SFLQG}. Probably, the key ingredient in overcoming the obstacle was the use of the scalar field. It helped to solve the long standing problem of derivation of a spin-foam model from the canonical theory but also opened the theory for applications. Having said this, we realize that it is not the link that closes the debate: the remaining problem is whether there is a precise relation between the canonical Loop Quantum Gravity and spin-foam models derived using the standard method by discretizing the theory first at the classical level and quantizing covariantly afterwards \cite{BC,EPRL,FK}. The first steps towards this direction would be to construct such spin-foam model of 4D Lorentzian gravity coupled to a scalar field, probably by coupling the EPRL/FK to such field. We hope that our derivation leaves some hints for such construction.

Our ideas straightforwardly generalize to irrotational dust and non-vanishing cosmological constant. Considering massive scalar field should not be much more involved (the biggest challenge in this case is that the mass term involves a product of the scalar field operators and the volume operator). We expect that considering other (polynomial) potentials may lead to a coupling of non-trivial Feynman diagrams for the matter part with the spin foams representing the gravitational degrees of freedom. For such potentials, the matter Hamiltonian should be split into the standard free part and interacting part. The free part should be treated together with the scalar field momentum and the Lorentzian part of the gravitational constraint while the interacting part should be treated together with the Euclidean part of the gravitational constraint.

Our expansion coincides with the expansion in a parameter defined by the Barbero-Immirzi parameter that has been recently proposed in \cite{TimeEvolution}. The authors use the perturbative expansion of the eigenvalues and the eigenvectors of the scalar constraint operator in order to find an approximate expression for the evolution operator. Our proposal provides a compact form of their expressions to all orders of the expansion and convenient graphical representation of the formulas in terms of spin foams. 

Let us notice that the model proposed in this paper is free from some divergence issues present in the spin-foam models of Quantum Gravity, for example in the EPRL/FK model \cite{EPRL,FK}. Firstly, the spins are fixed in all foams -- this eliminates the problem of sum over the spins. Secondly, although the model is Lorentzian, the structure group of the spin foams and spin networks is SU(2) -- this eliminates the problem with non-compactness of the SL(2,$\mathbb{C}$) group. Thanks to this property the expansion coefficients $A_M([s_\out],\varphi_\out(x_I);[s_\inn],\varphi_\inn(x_I))$ are finite. These two properties not only guarantee finitness of the expansion up to a finite order but also make the expansion coefficients much easier to compute, which hopefully will make it possible to perform numerical simulations of new quantum-gravitational physical phenomena using this model. Still there remains a problem of convergence of the series. Presently we study this issue in Loop Quantum Cosmology using numerical techniques. 

Let us underline that although the derivation of the spin-foam model that we performed in the full theory is based on the derivation in the reduced theory, they are not completely analogous. The difference lies in the splitting of the gravitational part of the constraint into $\hat{D}$ and $\hat{K}$ in LQC or $\CL[x]$ and $\CE[x]$ in LQG. In the first case $\hat{D}$ is the diagonal part of $\Cgr$ while $\hat{K}$ is its off-diagonal part in the volume eigenbasis. In the second case $\CL[x]$ and $\CE[x]$ are the Lorentzian and Euclidean parts of the constraint operator. One could also perform analogous splittings into Lorentzian and Euclidean parts in LQC. Our preliminary numerical research indicates that the convergence properties of the series with this splitting is worse than the one corresponding to the original splitting into $\hat{D}$ and $\hat{K}$.
\appendix
\renewcommand{\CE}[1][ ]{\hat{K}_{#1}}
\renewcommand{\eCE}[1][ ]{K_{#1}}
\renewcommand{\CL}[1][ ]{\hat{D}_{#1}}
\renewcommand{\eCL}[1][ ]{D_{#1}}
\section{Evaluation of the integral over $\ep$ by the contour method} \label{sc:Appendix_A}
\begin{figure}
\begin{center}
\includegraphics[scale=1.5]{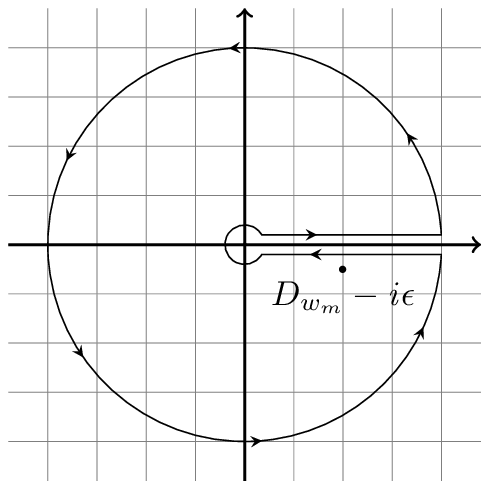}
\caption{Contour $\gamma$ in equation \eqref{eq:contour_integral}. The contour surrounds the poles $\eCL[w_m]\mp \iu \epsilon$. }
  \label{fig:contour}\end{center}
\end{figure}

We will perform the integral over $\ep$ in the physical scalar product $\phys{\nu_f,\phi_\out}{\nu_i,\phi_\inn}$ considered in Section \ref{sc:derivscfield} using the contour method. In a given order in the expansion in $\lambda$ the physical scalar product is given by a difference of two terms of the form:
\be
A_{\pm,\epsilon}(\nu_M,\ldots,\nu_0;\phi_\out,\phi_\inn)=\frac{1}{2\pi \iu}\int_{-\infty}^{\infty} d\ep\, 2\ep\, \theta(\ep) \frac{\eCE[\nu_M \nu_{M-1}]\ldots \eCE[\nu_1 \nu_{0}]}{\prod_{m=0}^{M}(\ep^2-\eCL[\nu_m]\pm \iu \epsilon)} e^{\iu \ep(\phi_\out-\phi_\inn)}.
\ee
 According to the notation introduced in Section \ref{sc:ACHHRVW} the order of the pole $\eCL[\nu_m]\mp \iu \epsilon$ is equal to $n_m$:
$$
A_{\pm,\epsilon}(\nu_M,\ldots,\nu_0;\phi_\out,\phi_\inn)=\eCE[\nu_M \nu_{M-1}]\ldots \eCE[\nu_1 \nu_{0}] \frac{1}{2\pi \iu}\int_{-\infty}^{\infty} d\ep\, 2\ep\, \theta(\ep) \frac{e^{\iu \ep(\phi_\out-\phi_\inn)}}{\prod_{m=0}^{p}(\ep^2-\eCL[w_m]\pm \iu \epsilon)^{n_m}} .
$$
Using the integral expression for the Heaviside theta:
$$
\theta(x)=-\frac{1}{2\pi \iu} \lim_{\epsilon\to 0^+} \int_{-\infty}^{\infty} d \tau\, \frac{1}{\tau+\iu \epsilon}\, e^{-\iu x \tau}
$$
we will write the integral as:
\begin{eqnarray}
\fl A_{\pm,\epsilon}(\nu_M,\ldots,\nu_0;\phi_\out,\phi_\inn)=\nonumber \\ \fl=\eCE[\nu_M \nu_{M-1}]\ldots \eCE[\nu_1 \nu_{0}] \frac{1}{2\pi^2}\lim_{\tilde{\epsilon}\to 0^+}\int_{-\infty}^{\infty} d\ep\,\int_{-\infty}^{\infty}\, d\tau\, \ep\, \frac{1}{\tau+\iu \tilde{\epsilon}} \frac{e^{\iu \ep(\phi_\out-\phi_\inn-\tau)}}{\prod_{m=0}^{p}(\ep^2-\eCL[w_m]\pm \iu \epsilon)^{n_m}}.\label{eq:amplitudepme}
\end{eqnarray}
The remaining problem is to calculate:
$$
I_{\pm,\epsilon}(\nu_M,\ldots,\nu_0;\phi_\out,\phi_\inn,\tau):= \int_{-\infty}^{\infty}\, d\ep\, \ep\, \frac{e^{\iu \ep(\phi_\out-\phi_\inn-\tau)}}{\prod_{m=0}^{p}(\ep^2-\eCL[w_m]\pm \iu \epsilon)^{n_m}},
$$
which we will do using the contour method. Let us denote by
$$
\Ep:= \ep^2
$$
and calculate a contour integral
\be\label{eq:contour_integral}
\sgn(\phi_\out-\phi_\inn+\tau)\frac{1}{2}\oint_\gamma\, d\Ep \frac{e^{\iu \sqrt{\Ep}|\phi_\out-\phi_\inn-\tau|}}{\prod_{m=0}^{p}(\Ep-\eCL[w_m]\pm \iu \epsilon)^{n_m}},
\ee
where the contour $\gamma$ is a keyhole contour depicted on the figure \ref{fig:contour} \footnote{It is contained in the lower sheet of the Riemann surface for square root ($0\leq\arg(\Ep)\leq 2\pi$)} and $|\phi_\out-\phi_\inn-\tau|$ denotes the absolute value of $\phi_\out-\phi_\inn-\tau$. Thanks to the absolute value the integral over the big circle vanishes as its radius tends to infinity. The integral over the small circle vanishes when its radius tends to zero. Therefore the contour integral tends to
\begin{eqnarray*}
\frac{1}{2}\oint_\gamma\, d\Ep \frac{e^{\iu \sqrt{\Ep}|\phi_\out-\phi_\inn-\tau|}}{\prod_{m=0}^{p}(\Ep-\eCL[w_m]\pm \iu \epsilon)^{n_m}}\to\\ \to \frac{1}{2}\int_0^\infty  d\Ep \frac{e^{\iu \sqrt{\Ep}|\phi_\out-\phi_\inn-\tau|}}{\prod_{m=0}^{p}(\Ep-\eCL[w_m]\pm \iu \epsilon)^{n_m}}-\frac{1}{2}\int_0^\infty  d\Ep \frac{e^{-\iu \sqrt{\Ep}|\phi_\out-\phi_\inn-\tau|}}{\prod_{m=0}^{p}(\Ep-\eCL[w_m]\pm \iu \epsilon)^{n_m}}=\\=\sgn(\phi_\out-\phi_\inn-\tau)\int_{-\infty}^{\infty}\, d\ep\, \ep\, \frac{e^{\iu \ep(\phi_\out-\phi_\inn-\tau)}}{\prod_{m=0}^{p}(\ep^2-\eCL[w_m]\pm \iu \epsilon)^{n_m}}.
\end{eqnarray*}
As a result,
$$
\sgn(\phi_\out-\phi_\inn-\tau)\frac{1}{2}\oint_\gamma\, d\Ep \frac{e^{\iu \sqrt{\Ep}|\phi_\out-\phi_\inn-\tau|}}{\prod_{m=0}^{p}(\Ep-\eCL[w_m]\pm \iu \epsilon)^{n_m}} \to \int_{-\infty}^{\infty}\, d\ep\, \ep\, \frac{e^{\iu \ep(\phi_\out-\phi_\inn-\tau)}}{\prod_{m=0}^{p}(\ep^2-\eCL[w_m]\pm \iu \epsilon)^{n_m}}.
$$
Using the residue theorem we obtain the expression for the integral $I_{\pm,\epsilon}(\nu_M,\ldots,\nu_0;\phi_\out,\phi_\inn,\tau)$:
\begin{eqnarray*}
I_{\pm,\epsilon}(\nu_M,\ldots,\nu_0;\phi_\out,\phi_\inn,\tau)=\\=\sgn(\phi_\out-\phi_\inn-\tau) \pi \iu \sum_{k=1}^p \Res(\frac{e^{\iu \sqrt{\Ep}|\phi_\out-\phi_\inn-\tau|}}{\prod_{m=0}^{p}(\Ep-\eCL[w_m]\pm \iu \epsilon)^{n_m}},\eCL[w_k]\mp \iu \epsilon)=\\=\sgn(\phi_\out-\phi_\inn-\tau) \pi \iu \sum_{k=1}^p \frac{1}{(n_k-1)!}\left. \frac{d^{n_k-1}}{d\Ep^{n_k-1}}\frac{e^{\iu \sqrt{\Ep}|\phi_\out-\phi_\inn-\tau|}}{\prod_{m\neq k}^p(\Ep-\eCL[w_{m}]\pm \iu \epsilon)^{n_{m}}}\right|_{\eCL[w_k]\mp \iu \epsilon}
\end{eqnarray*}
In the limit $\epsilon\to 0$ the expression becomes:
\begin{eqnarray*}
I_{\pm}(\nu_M,\ldots,\nu_0;\phi_\out,\phi_\inn,\tau):=\lim_{\epsilon\to 0}I_{\pm,\epsilon}(\nu_M,\ldots,\nu_0;\phi_\out,\phi_\inn,\tau)=\\=\sgn(\phi_\out-\phi_\inn-\tau) \pi \iu \sum_{k=1}^p \frac{1}{(n_k-1)!}\frac{d^{n_k-1}}{d\eCL[w_k]^{n_k-1}}\frac{e^{\mp\iu \sqrt{\eCL[w_k]}|\phi_\out-\phi_\inn-\tau|}}{\prod_{m\neq k}^p(\eCL[w_k]-\eCL[w_{m}])^{n_{m}}}.
\end{eqnarray*}
The physical scalar product \eqref{eq:scphysmat} depends on the difference of the two terms $I_{-}$ and $I_{+}$:
\begin{eqnarray*}
I_{-}(\nu_M,\ldots,\nu_0;\phi_\out,\phi_\inn,\tau)-I_{+}(\nu_M,\ldots,\nu_0;\phi_\out,\phi_\inn,\tau)=\\= \pi \iu \sum_{k=1}^p \frac{1}{(n_k-1)!}\frac{d^{n_k-1}}{d\eCL[w_k]^{n_k-1}}\frac{\sgn(\phi_\out-\phi_\inn-\tau)(e^{\iu \sqrt{\eCL[w_k]}|\phi_\out-\phi_\inn-\tau|}-e^{-\iu \sqrt{\eCL[w_k]}|\phi_\out-\phi_\inn-\tau|})}{\prod_{m\neq k}^p(\eCL[w_k]-\eCL[w_{m}])^{n_{m}}} =\\= \pi \iu \sum_{k=1}^p \frac{1}{(n_k-1)!}\frac{d^{n_k-1}}{d\eCL[w_k]^{n_k-1}}\frac{e^{\iu \sqrt{\eCL[w_k]}(\phi_\out-\phi_\inn-\tau)}-e^{-\iu \sqrt{\eCL[w_k]}(\phi_\out-\phi_\inn-\tau)}}{\prod_{m\neq k}^p(\eCL[w_k]-\eCL[w_{m}])^{n_{m}}},
\end{eqnarray*}
where the last equality follows from the fact that the sine is an odd function. 

\begin{figure}
\begin{center}
\includegraphics[scale=1.5]{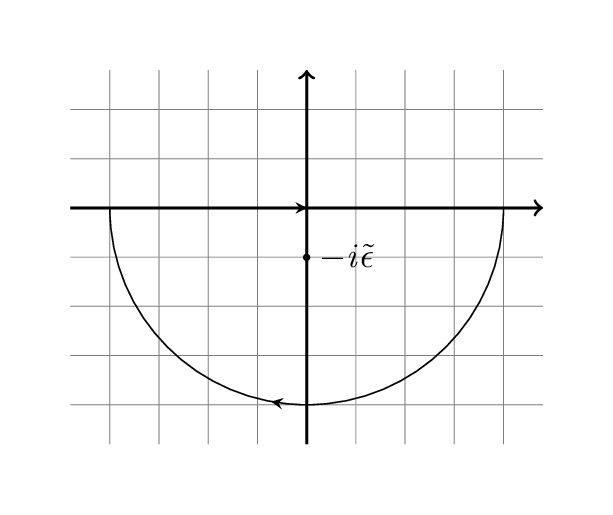}
\caption{The integral over $\tau$ in \eqref{eq:amplitudepme} is calculated using a sunset contour.}
\label{fig:contour_sunset}  \end{center}
\end{figure}

Let us calculate the integral over $\tau$ and take the limit $\tilde{\epsilon}\to 0$. Let us notice that $I_- - I_+$ as a function of $\tau$ has the form $P(\tau) e^{\iu \sqrt{\eCL[w_k]}\tau} + Q(\tau) e^{-\iu \sqrt{\eCL[w_k]} \tau}$, where $P$ and $Q$ are polynomials. The integral
$$
\frac{1}{2\pi^2}\lim_{\tilde{\epsilon}\to 0^+}\int_{-\infty}^{\infty} d\tau \frac{Q(\tau)}{\tau + \iu \tilde{\epsilon}} e^{-\iu \tau \sqrt{\eCL[w_k]}}
$$
can be calculated using the contour method with the sunset contour depicted on figure \ref{fig:contour_sunset} giving:
\begin{eqnarray*}
\frac{1}{2\pi^2}\lim_{\tilde{\epsilon}\to 0^+}\int_{-\infty}^{\infty} d\tau \frac{Q(\tau)}{\tau + \iu \tilde{\epsilon}} e^{-\iu \tau \sqrt{\eCL[w_k]}}=-\frac{2 \pi \iu}{2\pi^2} \lim_{\tilde{\epsilon}\to 0^+} Q(-\iu \tilde{\epsilon}) e^{-\tilde{\epsilon} \sqrt{\eCL[w_k]}}=\frac{ 1}{\pi\iu} Q(0)=\\= \sum_{k=1}^p \frac{1}{(n_k-1)!}\frac{d^{n_k-1}}{d\eCL[w_k]^{n_k-1}}\frac{e^{\iu \sqrt{\eCL[w_k]}(\phi_\out-\phi_\inn)}}{\prod_{m\neq k}^p(\eCL[w_k]-\eCL[w_{m}])^{n_{m}}}
\end{eqnarray*}
Similarly, by using similar sunset contour in the non-negative imaginary part region we show that:
$$
\lim_{\tilde{\epsilon}\to 0^+}\int_{-\infty}^{\infty} d\tau \frac{P(\tau)}{\tau + \iu \tilde{\epsilon}} e^{\iu \tau \sqrt{\eCL[w_k]}}=0.
$$

This allows us to express the physical scalar product as the following sum
$$
\phys{\nu_f,\phi_\out}{\nu_i,\phi_\inn}=\sum_{M=0}^{\infty} \lambda^M \sum_{{\nu_{M-1},\ldots, \nu_1 \atop \nu_m\neq \nu_{m+1}}} \eCE[\nu_M \nu_{M-1}]\ldots \eCE[\nu_1 \nu_{0}]\sum_{k=1}^p \frac{1}{(n_k-1)!}\frac{d^{n_k-1}}{d\eCL[w_k]^{n_k-1}}\frac{e^{\iu \sqrt{\eCL[w_k]}(\phi_\out-\phi_\inn)}}{\prod_{m\neq k}^p(\eCL[w_k]-\eCL[w_{m}])^{n_{m}}}.
$$
\bibliography{LQG}{}
\bibliographystyle{utcaps}
\end{document}